\documentclass[12pt]{article}
\usepackage{epsfig,amssymb}

\newcommand{\bq}{\begin{equation}}
\newcommand{\eq}{\end{equation}}
\newcommand{\bqa}{\begin{eqnarray}}
\newcommand{\eqa}{\end{eqnarray}}
\newcommand{\ben}{\begin{enumerate}}
\newcommand{\een}{\end{enumerate}}
\newcommand{\bc}{\begin{center}}
\newcommand{\ec}{\end{center}}
\newcommand{\bqb}{\begin{eqnarray*}}
\newcommand{\eqb}{\end{eqnarray*}}

\def\gsim{\gtrsim}

\def\pr#1#2#3{ Phys. Rev. ${\bf{#1}}$ (#2) #3}

\def\pl#1#2#3{ Phys. Lett. ${\bf{#1}}$ (#2) #3}
\def\prep#1#2#3{ Phys. Rep. ${\bf{#1}}$ (#2) #3}

\def\np#1#2#3{ Nucl. Phys. ${\bf{#1}}$ (#2) #3}
\def\zp#1#2#3{ Z. f. Phys. ${\bf{#1}}$ (#2) #3}
\def\epj#1#2#3{ Eur. Phys. J. ${\bf{#1}}$ (#2) #3}

\def\cpc#1#2#3{Comput. Phys. Commun. ${\bf{#1}}$ (#2) #3}

\def\eg{{\it e.g.\/}}

\def\etal{{\it et.al.\/}}

\global\nulldelimiterspace = 0pt

\def\L{ {\cal L }}

\def\sw{s_W}
\def\cw{c_W}

\def\mwd{m_W^2}
\def\mw{m_W}
\def\mz{m_Z}
\def\mzd{m_Z^2}

\def\t{\hat t}
\def\s{\hat s}
\def\u{\hat u}

\begin{document}
\pagenumbering{arabic}
\thispagestyle{empty}
\def\thefootnote{\fnsymbol{footnote}}
\setcounter{footnote}{1}

\begin{flushright}
PM/99-23 \\
THES-TP 99/05 \\
April 1999
 \end{flushright}
\vspace{2cm}
\begin{center}
{\Large\bf The $\gamma \gamma \to \gamma Z $ process at high energies
and the search for virtual SUSY effects.}\footnote{Partially
supported by the grant CRG 971470 and
by the Greek government grant PENED/95 K.A. 1795.}
 \vspace{1.5cm}  \\
{\large G.J. Gounaris$^a$, J. Layssac$^b$, P.I. Porfyriadis$^a$ and
F.M. Renard$^b$}\\
\vspace{0.7cm}
$^a$Department of Theoretical Physics, Aristotle
University of Thessaloniki,\\
Gr-54006, Thessaloniki, Greece.\\
\vspace{0.2cm}
$^b$Physique
Math\'{e}matique et Th\'{e}orique,
UMR 5825\\
Universit\'{e} Montpellier II,
 F-34095 Montpellier Cedex 5.\\
\vspace{0.2cm}

\vspace*{1cm}

{\bf Abstract}
\end{center}

We study the helicity amplitudes and the observables
of the process $\gamma \gamma \to \gamma Z$ at high energy.
As in the case of the $\gamma \gamma \to \gamma\gamma$ process
studied before, the relevant diagrams in the standard model (SM)
involve $W$, charged quark and lepton loops; while in SUSY we also have
contributions from  chargino, and charged sfermion or Higgs loop
diagrams. Above $250~ GeV$, the dominant SM amplitudes are
themselves dominated by the $W$ loop, and
as for $\gamma \gamma \to \gamma \gamma$, they  are
helicity conserving and almost purely imaginary.
We discuss the complementary   information provided
by $\gamma \gamma \to \gamma Z$ for the identification of possible non
standard effects. This process, together with
$\gamma \gamma \to \gamma \gamma $, should provide very
useful information on the nature of possible
 new physics particles, above the threshold of their
direct production.

\def\thefootnote{\arabic{footnote}}
\setcounter{footnote}{0}
\clearpage

\section{Introduction}

\hspace{0.7cm}In previous papers \cite{gggg, GPRgamma1} we studied
the process $\gamma \gamma \to\gamma \gamma $. The most striking
property of this process in the Standard Model (SM), is that
its whole  set of  possible
helicity amplitudes is strongly dominated by just the three
helicity conserving ones; which moreover
are almost purely imaginary. This simple property offers new
possibilities for improving the search for new physics (NP)
at high energy. Some of these possibilities
related to $\gamma \gamma \to\gamma \gamma $ have  been
investigated in the aforementioned papers.
We now extend this study to the process
$\gamma \gamma \to\gamma Z$.\par

In more detail, this  remarkable property of the
$\gamma \gamma \to \gamma \gamma $
processes is due to the fact that the Standard Model (SM)
amplitude first appears at the
1-loop level and at high energies it is dominated
by the $W$ loop contribution, which mainly enhances the imaginary
parts of the three helicity non flip amplitudes.
Thus in SM, this process is
dominated by just a few almost purely imaginary helicity
conserving amplitudes.   As we will see in the present
work, similar properties are also  valid for the process
$\gamma \gamma \to \gamma Z$ studied here. \par

This suggests
to use the $\gamma \gamma \to \gamma \gamma,~\gamma Z$
processes as a tool for searching for types of new physics
characterized by amplitudes with a substantial imaginary part that
can interfere with the SM one; like \eg~ effects
due to  chargino or charged slepton  loop diagrams above the threshold;
$s$-channel resonance production;
or new strong interactions inducing
unitarity saturating contributions to the NP amplitudes.\par

These studies could be achieved at a future $e^+e^-$ Linear
Collider (LC) \cite{LC} operating as a
$\gamma \gamma $
Collider ($LC_{\gamma \gamma}$) whose  c.m. energy
may be variable  and as high as $80\%$ of the initial $e^+e^-$
c.m. energy, by using the laser backscattering technique
\cite{LCgg, gamma97}. Polarized $\gamma \gamma$ beams can also be
obtained using initially polarized electron beams and lasers. \par

This search for NP through its virtual effects,
is complementary to the direct
production of new particles and it should help
identifying their nature; since it avoids the
model-dependent task of studying their decay modes, once they
are actually produced. More explicitly: the charged sparticle loop
contribution to $\gamma \gamma \to \gamma \gamma,~\gamma Z $,
is independent
of the many  parameters entering their decay modes and determining
\eg~ the soft SUSY breaking and the possible $R$-parity violating
sectors. \par

In the present paper  we study in detail the $\gamma \gamma \to
\gamma Z $ amplitudes in the standard and SUSY models. The
idea is to confirm and improve the searches for NP signatures
that can be done through direct production and the measurements of the
 $\gamma \gamma \to \gamma \gamma $ process. The situation for
such measurements should be more favorable in $\gamma \gamma \to
\gamma Z $  than in  $\gamma \gamma \to \gamma \gamma  $,
because the cross section is larger by about a factor 6.
Then, if a signal suspected in $\gamma \gamma \to \gamma \gamma $
is also seen here;  the detailed properties of the
SM departures, which now depend on the occurrence of the
$Z$ couplings, should allow some identification of the nature
of the effect. In particular  it should help identifying
the  $SU(2)\times U(1)$ quantum numbers of the
new physics particle contributing virtually. \par

In Sec.2 we discuss the main properties of the $W$, fermion
and scalar loop
contributions at high energies,  which had not been fully analyzed
before. This allows us to predict the type of effects expected in
case of New Physics (NP) contributions caused by new fermion or
scalar particle loops. We consider SUSY, as
an example of such an NP,  and we discuss the physical properties of
the contribution to the above amplitudes from a chargino or a charged
slepton,  showing how the presence of the $Z$ coupling can
distinguish them. For example the  magnitude of the contribution
changes notably when passing from a gaugino-like to a
higgsino-like contribution. And for a slepton loop,
even changes of sign appear,  when passing
from a slepton-L to a slepton-R case.\par

In Sec. 3, we study the $\gamma \gamma \to
\gamma Z $ cross sections in the standard and  SUSY models,
for various polarizations of the incoming photons. We identify the
sensitivity of these cross sections to various SUSY effects and we
discuss their observability in unpolarized and polarized $\gamma \gamma$
collisions, realized through the present ideas of laser backscattering.
Finally, in Sec. 4, we summarize the results and
give our conclusions.

The explicit expressions for the   $W$
\cite{Morgan, JikiaZ} and fermion loop \cite{GloverZ,  Morgan}
contributions to the helicity amplitudes are given in Appendix A,
using the non-linear gauge of
\cite{Dicus}.  We agree with the previous authors, apart from some
slight corrections affecting the  W contributions to some
small helicity amplitudes.
In addition,  we also give the  1-loop contribution
induced by a single charged scalar particle.
In Appendix B
simple asymptotic expressions for the helicity amplitudes are given,
which elucidate their physical properties at high energies.

\section{An overall view of the $\gamma \gamma \to \gamma Z $
amplitudes.}

\hspace{0.7cm}The invariant helicity amplitudes  $F_{\lambda_1 \lambda_2
\lambda_3\lambda_4}(\s,\t,\u)$ for the process $\gamma \gamma \to \gamma
Z $ with $\lambda_j$ being the helicities of the incoming
and outgoing particles,
are given in Appendix A. Altogether there are $3\times 2^3=24$
helicity amplitudes, which must of course satisfy the
constraints  from Bose (\ref{Bose1}).
In SM or  SUSY models,
charge conjugation enforces parity invariance at the 1-loop level,
which implies (\ref{parity}) and allows to express
all helicity amplitudes in terms of nine  analytic
functions; six for
transverse $Z$ and three for longitudinal $Z$:
 \bqa
& F_{++++}(\s,\t,\u)~ , ~ F_{+++-}(\s,\t,\u), &
  F_{++-+}(\s,\t,\u)~, ~ F_{++--}(\s,\t,\u)~, \nonumber \\
&F_{+-+-}(\s,\t,\u) = F_{+--+}(\s, \u,\t), &
F_{+---}(\s,\t,\u) = F_{+-++}(\s,\u,\t)~, \nonumber \\
& F_{+++0}(\s,\t,\u)~,~ F_{++-0}(\s,\t,\u) &
F_{+-+0}(\s,\t,\u) = F_{+--0}(\s,\u,\t) \ . \nonumber
\eqa
In Appendix A, we reproduce the\footnote{Certain
corrections are found for the W contributions to
some small helicity amplitudes,  when comparing
to  \cite{JikiaZ}.} $W$ and charged fermion
contributions of  \cite{JikiaZ, Morgan, GloverZ}, and we also
give the contributions to these amplitudes due to a scalar particle
loop. The essential difference between these amplitudes and those
of the $\gamma\gamma\to\gamma\gamma$ case presented
in \cite{gggg}, (apart from the obvious $\mz$ dependent
kinematic terms) is the appearance of
longitudinal $Z$ states,  and the replacement of the factor $Q_x^4$
by $Q_x^3g^Z_{Vx}$ for an $x$-particle loop contribution.\par

All results are given in terms of the standard 1-loop functions
$B_0$, $C_0$ and $D_0$, first introduced in \cite{Passarino}.
Explicit asymptotic expressions for these functions,
relevant for the $\gamma \gamma \to \gamma Z$ kinematics,
are given in Appendix B.  The dominant
W contributions to the corresponding
asymptotic helicity amplitudes are also given there; while the
corresponding expressions for the fermion and scalar
contributions  can be  easily   written using the
presented formulae.

\vspace{0.5cm}
 \underline{In the Standard Model} and for  $\s \gsim (250 GeV)^2 $,
the only non-negligible amplitudes in both the
$\gamma \gamma \to \gamma Z$ and
$\gamma \gamma \to \gamma \gamma $ process, are
$F_{\pm\pm\pm\pm}(\s,\t,\u)$ and
$F_{\pm\mp\pm\mp}(\s,\t,\u) = F_{\pm\mp\mp\pm}(\s,\u, \t)$,
which turn out to be completely dominated by the
$W$-loop contribution and
almost purely imaginary.  They satisfy
\bq
Im F^W_{\gamma\gamma\to\gamma Z}
\simeq {c_W\over s_W} Im F^W_{\gamma\gamma\to\gamma \gamma} \ \ .
\label{F-W-dom}
\eq
The fermion loop contribution to these same amplitudes
is much smaller, its real and imaginary parts are comparable
and roughly satisfies
\bq
F^f_{\gamma\gamma\to\gamma Z}
\simeq {g^Z_{Vf}\over Q_f} F^f_{\gamma\gamma\to\gamma \gamma}
\ \ , \label{F-fermion-dom}
\eq
where  for a standard quark or lepton $f$,
\bq
g^Z_{Vf} = {t^f_3 - 2Q_f s^2_W\over 2c_Ws_W} \ \ .
\eq
The rest of the amplitudes turn out to be much smaller
then the above dominant ones. For them, the real and imaginary parts
are roughly on the same footing; as well
the $W$- and fermion-loop contributions.

\begin{figure}[p]
\vspace*{-4cm}
\[
\epsfig{file=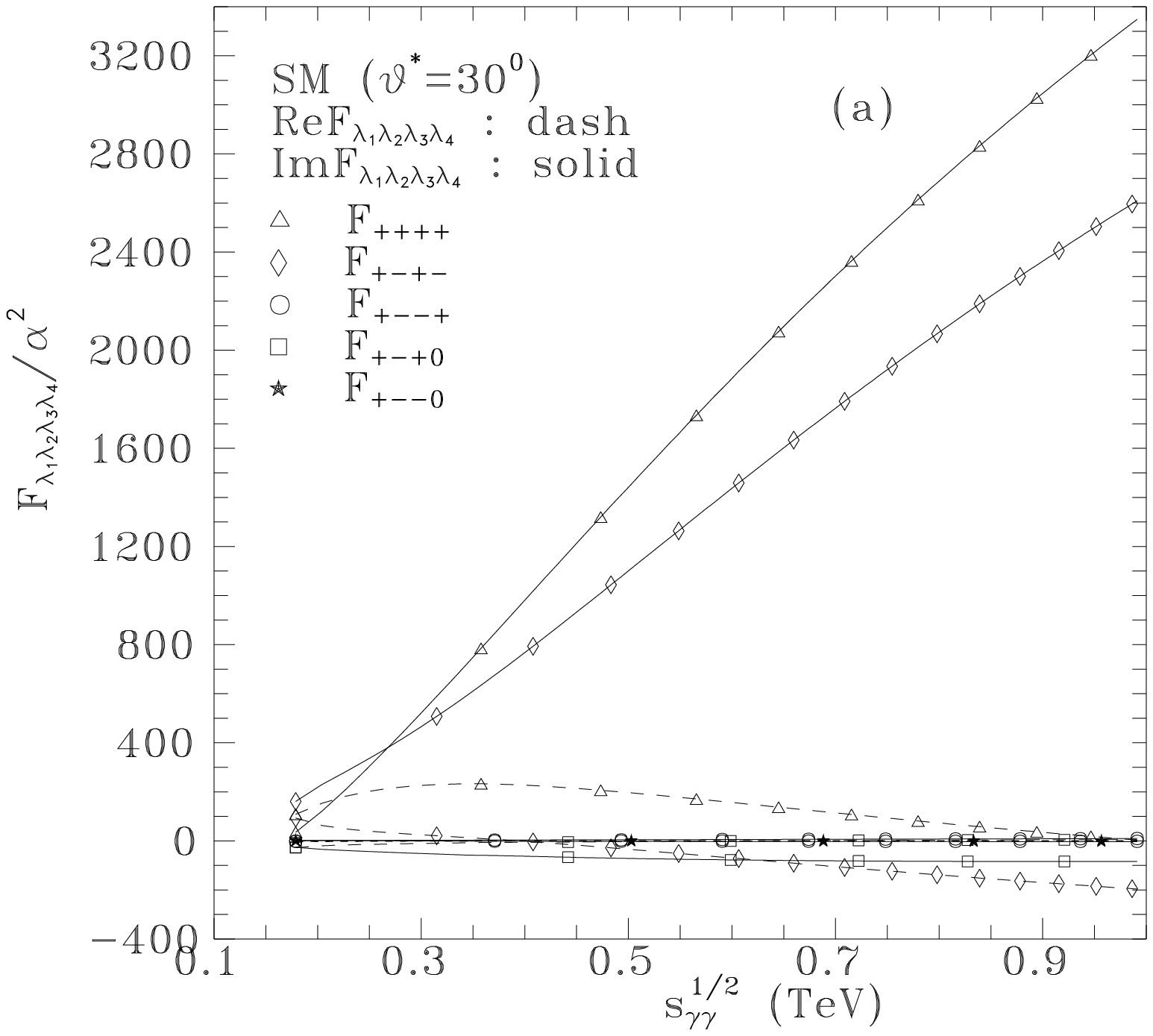,height=7.5cm}\hspace{0.5cm}
\epsfig{file=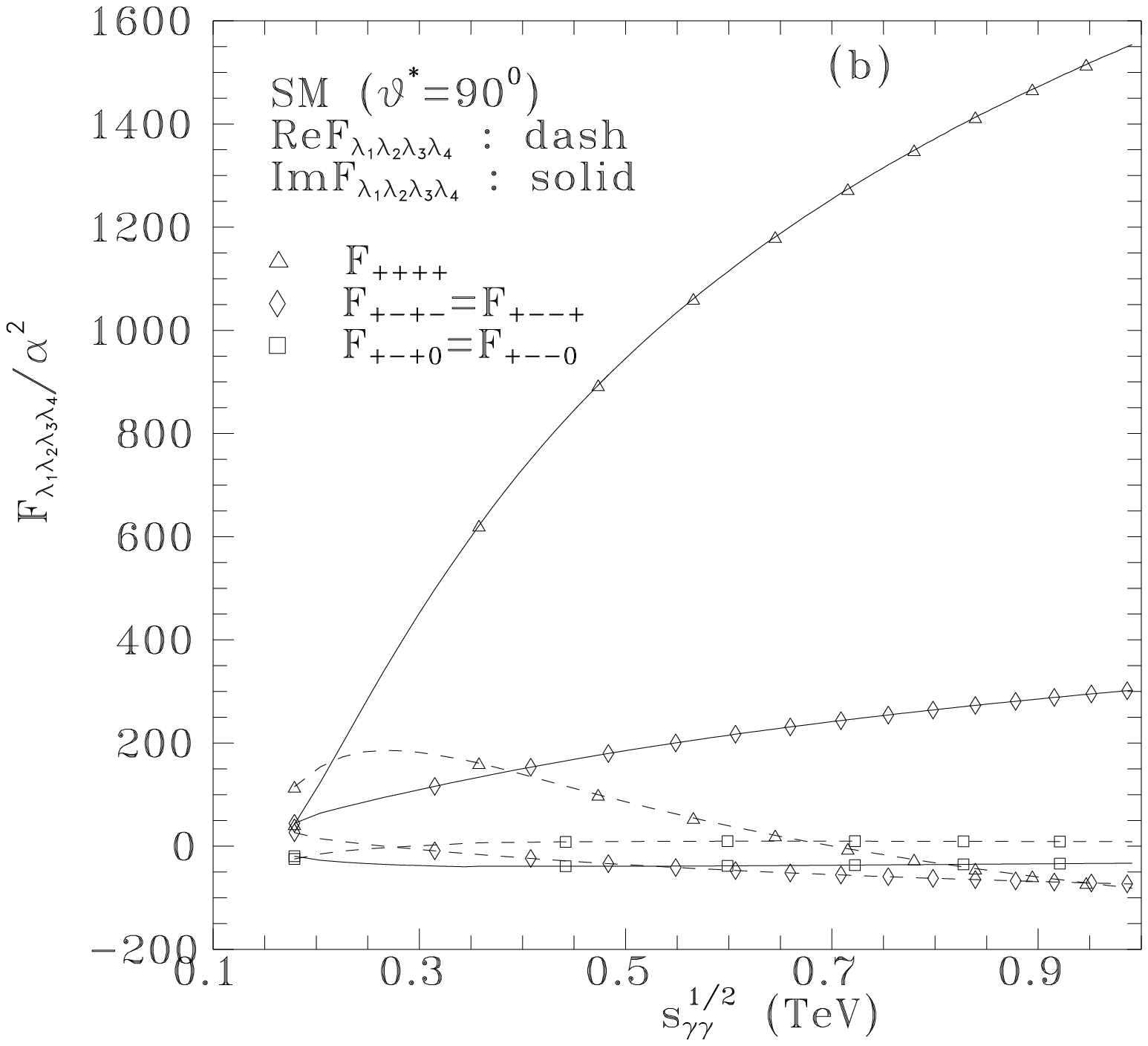,height=7.5cm}
\]
\vspace*{0.5cm}
\caption[1]{SM contribution to the dominant
$\gamma \gamma \to \gamma Z$ helicity amplitudes at
$\vartheta^*=30^0$ and $\vartheta^*=90^0$. All other amplitudes
are predicted to be smaller or about equal to $F_{+-+0}$ or
$F_{+--0}$.}
\label{sm-amp}
\end{figure}

Numerical results for these amplitudes using the exact 1-loop functions
are presented in Fig.\ref{sm-amp}a,b, and they agree with
the above expectations.
 Indeed  the real part of the large amplitudes
$F_{\pm\pm\pm\pm}(\s,\t,\u)$ and
$F_{\pm\mp\pm\mp}(\s,\t,\u) = F_{\pm\mp\mp\pm}(\s,\u, \t)$,
is always more than 4(15) times smaller than the imaginary part
at $\sqrt{\s} \simeq 0.3 (0.6) TeV $  .\par

As in the $\gamma \gamma \to \gamma \gamma $
case \cite{gggg}, the asymptotic  expressions in Appendix B
are quite accurate in describing  the large
SM helicity amplitudes $F_{\pm\pm\pm\pm}(\s,\t,\u)$
and $F_{\pm\mp\pm\mp}(\s,\t,\u) =
F_{\pm\mp\mp\pm}(\s,\u, \t)$, for the process
$\gamma \gamma \to \gamma Z $ also. This is due
to the fact that the double-log real contributions from
(\ref{Easym}, \ref{Ftildeasym}) always cancel out for physical
amplitudes, and the only
important contributions remaining are the single-log imaginary one.
For the rest of the helicity amplitudes, all log contributions either
cancel out or they are strongly suppressed by $\mwd/\s$ factors.\par

This confirms
 the fact that $\gamma \gamma \to \gamma Z $, much like
$\gamma \gamma \to \gamma \gamma $ scattering,  may  provide a
very useful tool for searching for types of New Physics (NP),
with largely  imaginary amplitudes \cite{GPRgamma1}. \par

\vspace{0.5cm}
\underline{We have thus computed the contributions of SUSY particles},
i.e. the contributions from a chargino  or a sfermion loop.\par

The contribution from the lightest positively charged chargino $\chi^+_1$
is obtained from the effective interaction (\ref{LZff}) by using
\cite{abdel}
\bq
g^Z_{V\chi}={1\over 2c_Ws_W}\left \{{3\over2}-2s^2_W +{1\over4}
[\cos(2\phi_L)+\cos(2\phi_R)]\right \} \ , \label{gVchiZ}
\eq
with
\bqa
\cos(2\phi_L)&=&-{M^2_2-\mu^2-2\mwd \cos(2\beta)\over
\sqrt{(M^2_2+\mu^2+2\mwd)^2-4[M_2\mu-\mwd \sin(2\beta)]^2}}
\ ,\nonumber\\
\cos(2\phi_R)&=&-{M^2_2-\mu^2+2\mwd \cos(2\beta)\over
\sqrt{(M^2_2+\mu^2+2\mwd)^2-4[M_2\mu-\mwd \sin(2\beta)]^2}}
\ ,
\eqa
and
\bq
M^2_{\chi_1^+}={1\over2}\{M^2_2+\mu^2+2\mwd -
\sqrt{(M^2_2+\mu^2+2\mwd )^2-4[M_2\mu-\mwd \sin(2\beta)]^2}~\}.
\label{chimass}
\eq

Using the formulae (\ref{f++++} -\ref{f+-+0})
in Appendix A, together  with
the exact 1-loop calculation from  \cite{Oldenborgh},
and (\ref{gVchiZ}, \ref{chimass});
 we present in Fig.\ref{chargino-amp} the results for
two almost "extreme" situations corresponding
to a light chargino of mass $M_{\chi_1^+}\simeq
95~GeV$ and  $\tan\beta=2$. In the first case the chargino nature
is taken gaugino-like, by choosing (see Fig.\ref{chargino-amp}a,b)
\bq
M_2=0.081~TeV~~~~,~~~~ \mu=-0.215~TeV~~~~,~~~~~~g^Z_{V\chi}=
1.72 \  \label{gaugino-par};
\eq
while in the second case it is taken "higgsino-like" by choosing
(see Fig.\ref{chargino-amp}c,d)
\bq
M_2=0.215~TeV~~~~,~~~~ \mu=-0.081~TeV~~~~~,~~~~~g^Z_{V\chi}
= 0.73 \ . \label{higgsino-par}
\eq \par

We also consider the two L,R-slepton cases,  with
$M_{\tilde l}=0.1~TeV$, $Q_{\tilde l}=-1$ and
\bq
g^Z_{V\tilde l}={1\over c_Ws_W}[t^{\tilde l}_3 - Q_{\tilde l} s^2_W]
\ \ .
\eq
\noindent
For  $t^{\tilde l}_3=-{1\over2}$, these lead to
$g^Z_{V\tilde l}=-0.65$  (case L); while for
$t^{\tilde l}_3=0$ we find  $g^Z_{V\tilde l}=+ 0.54$ (case R).
As a result, a change of sign appears between $L$
and $R$ slepton contributions.
The corresponding
results for a slepton, are derived using (\ref{S++++} -
\ref{S+-+0}) and presented in
Fig.\ref{slepton-amp}a-d.

As seen for both cases of Fig.\ref{chargino-amp}a-d
and Fig.\ref{slepton-amp}a-d,  the
real and imaginary parts of the fermion  or scalar loop
contributions to the $\gamma \gamma \to \gamma Z$ amplitudes
above threshold, are more or less   on the same footing.
It is also seen that
immediately  above the threshold, an imaginary
contribution to the  $F_{\pm\pm\pm\pm}(\s,\t,\u)$ and
$F_{\pm\mp\pm\mp}(\s,\t,\u)=F_{\pm\mp\mp\pm}(\s,\u, \t)$
amplitude starts developing,
which can interfere with the SM one and produce a measurable effect.
The slepton contribution is smaller than the chargino one,
by about a factor of seven though.\par

As compared to the $\gamma\gamma\to\gamma\gamma$ case, we also
notice the presence of large longitudinal $Z$
amplitudes, for both the chargino and the slepton cases.
However these are not easily observable, since they do  not
find a corresponding large longitudinal SM amplitude
to interfere with. So at the end,
 they will produce very small effect.

For  transverse amplitudes, as a consequence of the
$g^Z_{Vf}/Q_f$ factor, the effect in the
$\gamma \gamma \to \gamma Z$ case is larger (weaker) than the
$\gamma\gamma\to\gamma\gamma$ effect in the gaugino (higgsino)
cases. The corresponding effect in the slepton cases,
changes sign when
passing from the L-slepton to the R-slepton contribution. These
properties will directly reflect in the threshold effects
appearing in the cross sections that we study in the next section.

\section{ The $\gamma \gamma \to \gamma Z$ Cross sections}

\hspace{0.7cm}We next explore
the possibility to use polarized or unpolarized
$\gamma\gamma$ collisions in an LC operated in the $\gamma
\gamma $ mode,  through laser backscattering and the procedure described
in  \cite{Tsi, GPRgamma1}.  The assumption of
Parity invariance leads to the following form for the
$\gamma \gamma \to \gamma Z $
cross section (note the factor 2 as compared to the
$\gamma \gamma \to \gamma \gamma $ due to the presence of
a non symmetric final state)
\bqa
{d\sigma\over d\tau d\cos\vartheta^*}&=&{d \bar L_{\gamma\gamma}\over
d\tau} \Bigg \{
{d\bar{\sigma}_0\over d\cos\vartheta^*}
+\langle \xi_2 \xi_2^\prime
\rangle{d\bar{\sigma}_{22}\over d\cos\vartheta^*}
+\langle\xi_3\rangle \cos2\phi \,
{d\bar{\sigma}_{3}\over d\cos\vartheta^*}
+\langle\xi_3^ \prime\rangle\cos2\phi^\prime
{d\bar\sigma_3^\prime\over d\cos\vartheta^*}
\nonumber\\
&&+\langle\xi_3 \xi_3^\prime\rangle
\left[{d\bar{\sigma}_{33}\over d\cos\vartheta^*}
\cos2(\phi+\phi^\prime)
+{d\bar{\sigma}^\prime_{33}\over
d\cos\vartheta^*}\cos2(\phi- \phi^\prime)\right ]\nonumber\\
&&+ \langle\xi_2 \xi_3^\prime\rangle\sin2 \phi^\prime
{d\bar{\sigma}_{23}\over d\cos\vartheta^*}-
\langle\xi_3 \xi_2^\prime\rangle\sin2\phi\,
{d\bar{\sigma}_{23}^\prime\over d\cos\vartheta^*} \Bigg \} \ \ ,
\label{sigpol}
\eqa
where
\bqa
{d\bar \sigma_0\over d\cos\vartheta^*}&=&
\left ({\beta_Z\over64\pi\hat{s}}\right )
\sum_{\lambda_3\lambda_4} [|F_{++\lambda_3\lambda_4}|^2
+|F_{+-\lambda_3\lambda_4}|^2] ~ ,  \label{sig0} \\
{d\bar{\sigma}_{22}\over d\cos\vartheta^*} &=&
\left ({\beta_Z\over64\pi\hat{s}}\right )\sum_{\lambda_3\lambda_4}
[|F_{++\lambda_3\lambda_4}|^2
-|F_{+-\lambda_3\lambda_4}|^2]  \ , \label{sig22} \\
{d\bar{\sigma}_{3}\over d\cos\vartheta^*} &=&
\left ({-\beta_Z\over32\pi\hat{s}}\right ) \sum_{\lambda_3\lambda_4}
Re[F_{++\lambda_3\lambda_4}F^*_{-+\lambda_3\lambda_4}]  \ ,
\label{sig3} \\
{d\bar{\sigma}_{3}^\prime \over d\cos\vartheta^*} &=&
\left ({-\beta_Z\over32\pi\hat{s}}\right ) \sum_{\lambda_3\lambda_4}
Re[F_{++\lambda_3\lambda_4}F^*_{+-\lambda_3\lambda_4}]  \ ,
\label{sig3p} \\
{d\bar \sigma_{33} \over d\cos\vartheta^*}& = &
\left ({\beta_Z\over64\pi\hat{s}}\right ) \sum_{\lambda_3\lambda_4}
Re[F_{+-\lambda_3\lambda_4}F^*_{-+\lambda_3\lambda_4}] \ ,
\label{sig33} \\
{d\bar{\sigma}^\prime_{33}\over d\cos\vartheta^*} &=&
\left ({\beta_Z\over64\pi\hat{s}}\right ) \sum_{\lambda_3\lambda_4}
Re[F_{++\lambda_3\lambda_4}F^*_{--\lambda_3\lambda_4}] \  ,
\label{sig33prime} \\
{d\bar{\sigma}_{23}\over d\cos\vartheta^*}& = &
\left ({\beta_Z\over 32\pi\hat{s}}\right ) \sum_{\lambda_3\lambda_4}
Im[F_{++\lambda_3\lambda_4}F^*_{+-\lambda_3\lambda_4}] \ ,
\label{sig23}\\
{d\bar{\sigma}_{23}^\prime \over d\cos\vartheta^*}& = &
\left ({\beta_Z\over 32 \pi\hat{s}}\right ) \sum_{\lambda_3\lambda_4}
Im[F_{++\lambda_3\lambda_4}F^*_{-+\lambda_3\lambda_4}] \ ,
\label{sig23p}
\eqa
are expressed in terms of the $\gamma \gamma \to \gamma Z$
amplitudes given in Appendix A. In (\ref{sig0} - \ref{sig23p}),
 $\beta_Z=1-\mzd/\s$ is the $Z$ velocity in the $\gamma \gamma $
rest frame, while $\vartheta^*$ is the scattering angle, and
$\tau \equiv s_{\gamma \gamma}/s_{ee}$.
Note that $d\bar \sigma_0/ d\cos\vartheta^*$ is the unpolarized
cross section and it is the only $\bar \sigma_j$ quantity which is
positive definite.  We also note that
$d\bar \sigma_j/ d\cos\vartheta^*$ are forward-backward symmetric
except those below which satisfy
\bqa
{d\bar \sigma_3^\prime  \over d\cos\vartheta^*}
\Bigg\vert_{\vartheta^*}
& = & {d\bar \sigma_3 \over d\cos\vartheta^*}
\Bigg\vert_{\pi-\vartheta^*} \  \ , \label{3-3p}\\
{d\bar \sigma_{23}^\prime \over d\cos\vartheta^*}
\Bigg\vert_{\vartheta^*}& = &
{d\bar \sigma_{23}\over d\cos\vartheta^*}
\Bigg\vert_{\pi-\vartheta^*}
\ \ \label{23-23p}.
\eqa \par

The results for the cross sections $\bar \sigma_j$, integrated in the
range $30^0 \leq \vartheta^* \leq 150^0$,  are given in
Fig.\ref{sig}a-f, for the standard model (SM);
as well as for the   cases  of including the contributions from
a single chargino or a single charged slepton with mass of about
 95 or 100 GeV respectively.\par

As seen in Fig\ref{sig}a, the effects in the unpolarized cross
section $\bar \sigma_0$ are
consistent with those  expected  from the dominant
imaginary amplitudes quoted in the previous section. In more detail,
the (gaugino, higgsino) effects are of the order of (10, 5)\%,
respectively; while the slepton ones are an order of magnitude
smaller.\par

The \underline{relative} (NP versus SM) effects
are somewhat reduced in $\bar \sigma_{22}$, but they are
largely enhanced in the other $\bar \sigma_{j}$ cross
sections; (compare Fig.\ref{sig}b with Fig.\ref{sig}c-f.)
However most of these later cross sections have small absolute
values, making their observation doubtful. Only $\bar \sigma_{3}$
and $\bar \sigma_{33}$ are of the order of a few $fb$. The SUSY
effects are here notably enhanced as compared to the SM contributions,
reaching  the 25\% level in some cases. Particularly striking is
the SUSY effect for $\bar \sigma_{33}$ and
$\bar \sigma_{33}^\prime$ near  but above threshold.
 For $\bar \sigma_{33}^\prime$, such a behaviour
also occurred in the $\gamma\gamma\to\gamma\gamma$ case
\cite{gggg}, and it may be useful for  disentangling
of the various SUSY examples.\par

In the present case, the option to select longitudinal polarization
for the outgoing $Z$, is also available. In  this case
remarkable SUSY effects, strongly dependent on the type of particle
running along the loop, are generated.
However, the absolute values
of the related cross sections are unfortunately less
than 1 $fb$; rendering these effects unobservable. See
Fig.\ref{long}a-c where $ \bar \sigma_{0}(Z_L)$,
$\bar \sigma_{22}(Z_L)$ and $\bar \sigma_{33}(Z_L)$ are
presented; (the  other ones are less than $0.1~fb$.) \par

The angular distributions of the various
$d \bar \sigma_j/d\cos \vartheta^*$ are illustrated for
$\sqrt{s}=0.4~TeV$ in Fig.\ref{ang}a-f.
Restricting the discussion to
those  "cross sections" whose absolute values are larger than
$1~fb$; one observes that forward-backward peaks arise only for
$d \bar \sigma_0/d\cos \vartheta^*$.
Among the rest, the most interesting ones
are $d\bar \sigma_{22}/ d\cos\vartheta^*$,
$d\bar \sigma_{3} / d\cos\vartheta^*$ (note (\ref{3-3p})) and
$d\bar \sigma_{33}/ d\cos\vartheta^*$; which in fact have a
forward deep. At a weaker level,
a similar result is also true for $\gamma\gamma\to\gamma\gamma$,
where of course all "cross sections" are forward-backward symmetric;
(but no figures are shown  in \cite{gggg}). Fig.\ref{ang}a-f also
show that SUSY effects often appear mostly pronounced at large
angles.\par

To get a feeling on  the observability  of the various quantities
$\bar \sigma_j$ appearing in (\ref{sigpol}), we
next turn to the experimental aspects of the $\gamma \gamma$ collision
realized through laser backscattering \cite{LC, LCgg}.
 The quantity $d\bar L_{\gamma\gamma}/d\tau$
in  (\ref{sigpol}), describes the
photon-photon luminosity
per unit $e^-e^+$ flux, in an LC operated in the $\gamma \gamma$ mode
\cite{LCgg}.
The Stokes parameters $\xi_2$, $\xi_3$ and the polarization angle
$\phi$ in (\ref{sigpol}), determine the normalized
helicity density matrix
of one of the backscattered photons
$\rho^{BN}_{\lambda \tilde \lambda}$
through the formalism described in Appendix B of \cite{gggg};
compare eq.(B4) of \cite{gggg} and \cite{Tsi}.
The corresponding parameters for the other backscattered photon are
denoted by a prime. The numerical expectations for
 $d\bar L_{\gamma\gamma}/d\tau$, $\langle \xi _j \rangle $,
$\langle \xi _j^\prime \rangle $ and
$\langle \xi_i \xi _j^\prime \rangle $ are given
in Appendix B and Fig.4 of \cite{gggg}.
To estimate the expected number of events,
one should  multiply the
cross sections in (\ref{sigpol}) by the
$e^+e^-$ luminosity $\L_{ee}$, whose  presently contemplated value
for the LC project is $\L_{ee} \simeq 500~-~1000~
fb^{-1}$ per one or two years of running in \eg\@ the high
luminosity TESLA mode at energies of $350-800~ GeV$  \cite{LC}.\par

We next turn to  the expected statistical accuracies
for the various $\bar \sigma_j$. The
 relative uncertainty for  the unpolarized cross
section $[ \bar \sigma_0(\langle \cos\vartheta^* \rangle)]$,
calculated by integrating the respective differential cross section
over a certain reduced
energy bin $\Delta\tau$ and an angular bin
$\Delta \cos\vartheta^*$,  is:
\bq
{\delta[ \bar \sigma_0(\langle \cos\vartheta^* \rangle ] \over
[ \bar \sigma_0(\langle \cos\vartheta^* \rangle]  }=
[ \L_{ee}(\Delta\tau)
( \Delta \cos\vartheta^*)( {d \bar L_{\gamma\gamma}
\over d\tau})({d\bar \sigma_0\over d\cos\vartheta^*})]^{-{1\over2}}
\ \ . \label{accsig}
\eq
For the other "cross sections", for whose measurement
we need various combinations of longitudinal $e^\pm$, and
longitudinal and transverse laser polarizations, we perform an
analysis similar to the one in Section 3 of  \cite{gggg}.
For simplicity, we define
\bq
 R_j(\langle \cos\vartheta^*\rangle )\equiv
{ [ \bar \sigma_{ij}(\langle \cos\vartheta^* \rangle)] \over
[ \bar \sigma_0(\langle \cos\vartheta^* \rangle)]_{SM} } \ \ .
\label{Rj}
\eq
Then,   the absolute uncertainties
$\delta [ \bar \sigma_{ij} (\langle \cos\vartheta^* \rangle)]$
satisfy
\bq
\delta R_j(\langle \cos\vartheta^*\rangle)
={\delta [ \bar \sigma_{ij}(\langle \cos\vartheta^* \rangle) ] \over
[ \bar \sigma_0(\langle \cos\vartheta^* \rangle)]_{SM} }
={1\over c_j}
[ \L_{ee}(\Delta\tau)( \Delta \cos\vartheta^*)( {d \bar L_{\gamma\gamma}
\over d\tau})({d\bar \sigma_0\over d\cos\vartheta^*})]^{-{1\over2}}
\label{accR} \ \ ,\eq
\noindent
where $ c_j= \sqrt{2} \langle \xi_2 \xi_2^\prime \rangle$,
 $\sqrt{2} \langle \xi_3 \rangle$,
$ \langle \xi_3 \xi_3^\prime \rangle$,
$\sqrt{2}\langle \xi_2 \xi_3^\prime \rangle$, for
$R_{22}$, $R_{3}$, ($R_{33}$ or $R'_{33}$), ($R_{23}$ or
$R'_{23}$) respectively.\par

To estimate these, we take the numerical
values for the photon spectra and polarization degrees given
in Appendix B and Fig.4 of \cite{gggg}. For the
$e^+e^-$ luminosity we assume  $1000~fb^{-1}$.
Using  then bins of the order
of $\Delta\tau\simeq 0.4$, $\Delta \cos\vartheta^*\simeq 1$, and
$d \bar L_{\gamma\gamma}/ d\tau \gsim 1 $, as well as
an unpolarized differential cross section of the order
of $30~fb$ (see Fig.\ref{ang}a); one obtains a relative uncertainty
of the order of 1\%, for the unpolarized cross section in
(\ref{accsig}).

For the ratios $R_j$ defined in (\ref{Rj}),
the factor $1/ c_j$ will increase
the absolute uncertainty in (\ref{accR}).
According to Fig.4,5  of \cite{gggg},
this factor depends strongly on the backscattering configurations
and on the reduced energy range. It can easily vary between
1 and 10. But if the  kinematic range to  be studied
is known, then the backscattering configuration can be tuned
to optimize the flux spectrum. Thus, for the time being,
we can roughly conclude that the accuracy at which the ratios $R_j$
can be measured, should lie between 1 and 10\%. This means
that it is reasonable to expect an absolute uncertainty of
the order of $0.3~fb$ for
$d\bar{\sigma}_{0}/ d\cos\vartheta^*$ at large angles;
and something in the range  $(0.3 ~ - ~3)~fb$, for  the other
$d\bar{\sigma}_{ij}/ d\cos\vartheta^*$.\par

These values have to be compared with the NP effects expected
on the corresponding observables. Thus, from Fig.4,6 one sees that
the unpolarized integrated cross section should very sensitive
to chargino effects, a sensitivity  characterized by
a statistical significance notably
increased as compared with the $\gamma\gamma\to\gamma\gamma$
case; (up to 10 SD instead of 3 SD if the chargino
is in the 100 GeV mass range).
For  slepton searches, the
situation in $\gamma \gamma \to \gamma Z$
is similar to the one  in $\gamma \gamma \to \gamma \gamma$,
 because of the small $Z$-slepton
couplings.\par

The illustrations  given in the present paper are for a chargino or
slepton in 100GeV mass range.
For higher masses, the relative merits of  the
$\gamma\gamma\to\gamma Z$  and
$\gamma\gamma\to\gamma\gamma$ processes\footnote{ In \cite{gggg}
we gave some illustration for sparticles at $250~GeV$ in the
$\gamma\gamma\to\gamma\gamma$ case.} remain about the same.
We expect therefore that $\gamma\gamma\to\gamma Z$ should
be very helpful for sparticle searches with mass up to 300 GeV.
As a final remark, we recall that in
$\gamma\gamma\to\gamma\gamma$,
if several SUSY particles exist within a given mass range,
then their effects are all positive and cumulate in $\bar \sigma_0$.
This is not necessarily the case in $\gamma\gamma\to\gamma Z$, because
the $g^Z_{Vx}$ can have different signs as we have seen in Sect.2.

\section{Conclusions}

\hspace{0.7cm}In this paper we have
extended our previous analysis of the
helicity amplitudes and observables
in the process $\gamma \gamma \to \gamma
\gamma$ at high energies, to the $\gamma \gamma \to \gamma Z$ case.\par

It appears that both processes share the
spectacular property that in the Standard Model
and at  energies above $0.25 ~TeV$,
only three independent helicity conserving amplitudes are
important, which moreover are almost purely imaginary.
Exactly as it would had been predicted about 30 years ago,
 on the
basis of Vector Meson Dominance and Pomeron exchange!
These three amplitudes are $F_{\pm\pm\pm\pm}(\s,\t,\u)$ and
$F_{\pm\mp\pm\mp}(\s,\t,\u) = F_{\pm\mp\mp\pm}(\s,\u,\t)$.
Thus the $\gamma \gamma \to \gamma \gamma$
and $\gamma \gamma \to \gamma Z$
processes should be   excellent tools
for searching  for virtual new physics contributions
characterized by important imaginary contributions.
This means that they should very helpful in identifying the
nature of nearby new particles, which can also be directly excited.
But they should not be of much use for studying high scale
NP effects described by effective lagrangians, which naturally
lead to real amplitudes.\par

This has been illustrated for the particular SUSY cases of a
single chargino or  charged slepton contribution.
Clear threshold effects in the various observables appear due to the
interference of the imaginary parts of the SUSY amplitudes with the SM
ones. These
contributions depend of course on the mass and quantum numbers
of the SUSY partners, but are independent of the many model-dependent
parameters entering their decay modes, contrary to the case of direct
SUSY particle production. Thus, the study of the
$\gamma \gamma \to \gamma \gamma,~\gamma Z$ cross sections should offer
complementary  information, to the one obtained from direct
SUSY production cross sections.
We have indeed found that the unpolarized
$\gamma \gamma \to \gamma \gamma,~\gamma Z $ cross sections
$\bar \sigma_0$, are
most sensitive to a chargino  loop contribution.
For a light chargino the
signal is at most of 3 to 4 SD in $\gamma \gamma \to
\gamma \gamma$, while it can reach 10 SD in the
$\gamma \gamma \to \gamma Z $ case.
 For a single
charged slepton with a $100~GeV$ mass, we have found that the
corresponding effect on $\bar
\sigma_0$ is an order of magnitude smaller.
Angular distributions are most sensitive at large angles
($|\cos\theta^*|<0.5$) in both cases.
Polarization should allow to test the nature of the particles involved
in the loop. There are eight different observables in
$\gamma \gamma \to \gamma Z $ and six in $\gamma \gamma \to
\gamma\gamma $. Only five of them ($\bar \sigma_0$,
$ \bar \sigma_{22}$, $\bar \sigma_3$,
$ \bar \sigma_{33}$, $ \bar \sigma_{33}^\prime$),
will be measurable
with sufficient accuracy  to allow checks of the global picture.
This requires an optimization of the laser backscattering
procedure though.\par

The comparison of the situations in $\gamma \gamma \to
\gamma\gamma $ and  $\gamma \gamma \to \gamma Z $ is
very instructive.
It is first important to notice,
that in $\gamma \gamma \to \gamma\gamma $,
both, the charged fermion and the
charged scalar particle loops, {\it increase } the SM
prediction for $\bar \sigma_0$. If, as seems quite plausible,
a chargino, as well as all six charged sleptons  and $\tilde
t_1$, lie in the (100-250)GeV  mass range; then
  a clear signal could be seen in
$\bar \sigma _0(\gamma \gamma \to \gamma\gamma  )$.
The study of $\gamma \gamma \to \gamma Z $, including also
polarization effects, should then give information
on  the origin of the signal.\par

Similar type of effects could also  appear for  other virtual
NP contributions of fermionic or scalar nature; like \eg\@
heavy fermions,  technifermions, charged Higgses,  pseudogoldstone
bosons or even heavy charged vector bosons.
In the $\gamma \gamma \to \gamma \gamma$ case, the effect is only
controlled by the electric charge; while in
$\gamma \gamma \to \gamma Z$, the $g^Z_{Vx}$ coupling also enters.
There exist other process of this type, namely
$\gamma \gamma \to ZZ$ and   $\gamma \gamma \to HH,~ HZ$,
which only receive SM contributions at 1-loop, and could be equally
interesting for NP searches. However they are essentially controlled
by the properties of the Higgs sector and deserve separate studies
which are in progress.\par

In any case it appears to us that
$\gamma \gamma \to \gamma \gamma,~\gamma Z$ are very clean
processes which
should supply excellent tools for NP searches, and should add to
the interest in providing for the eventual realization of
the $\gamma \gamma$  mode in the  high energy LC colliders.\par

\newpage

\renewcommand{\theequation}{A.\arabic{equation}}
\renewcommand{\thesection}{A.\arabic{section}}
\setcounter{equation}{0}
\setcounter{section}{0}

{\large \bf Appendix A: The $\gamma \gamma \to \gamma Z $
amplitudes in SM and SUSY.}

The invariant helicity amplitudes for  the process
\bq \gamma (p_1,\lambda_1) \gamma (p_2,\lambda_2) \to \gamma
(p_3,\lambda_3) Z (p_4,\lambda_4) \ \ , \label{gggg-process}
\eq are denoted as\footnote{Their sign is related to the sign of
the $S$-matrix through   $S_{\lambda_1 \lambda_2
\lambda_3\lambda_4}= 1+i (2\pi)^4 \delta(p_f-p_i)
F_{\lambda_1 \lambda_2 \lambda_3\lambda_4}$.  We use
the Jacob-Wick convention.}
$F_{\lambda_1
\lambda_2 \lambda_3\lambda_4}(\s,\t,\u)$, where the momenta and
helicities of the incoming and outgoing photons are indicated in
parenthesis, and $\s=(p_1+p_2)^2$, $\t=(p_1-p_3)^2$,
$\u=(p_1-p_4)^2$. Bose statistics demands
\bqa
F_{\lambda_1 \lambda_2 \lambda_3\lambda_4}(\s,\t,\u) &=&
F_{\lambda_2 \lambda_1 \lambda_3\lambda_4}(\s,\u,\t)
(-1)^{1-\lambda_4} \ ,
\label{Bose1}
\eqa
while, if parity invariance also holds, we get
the additional constraint
\bqa
F_{\lambda_1 \lambda_2 \lambda_3\lambda_4}(\s,\t,\u) &=&
F_{-\lambda_1-\lambda_2- \lambda_3-\lambda_4}(\s,\t,\u)
(-1)^{1-\lambda_4} \  .
 \label{parity}
\eqa

As a result, the 24  helicity amplitudes
may be expressed in terms
of just the nine amplitudes
\bqa
& F_{++++}(\s,\t,\u)~ , ~ F_{+++-}(\s,\t,\u), &
  F_{++-+}(\s,\t,\u)~, ~ F_{++--}(\s,\t,\u)~, \nonumber \\
&F_{+-+-}(\s,\t,\u) = F_{+--+}(\s, \u,\t), &
F_{+---}(\s,\t,\u) = F_{+-++}(\s,\u,\t)~, \nonumber \\
& F_{+++0}(\s,\t,\u)~,~ F_{++-0}(\s,\t,\u) &
F_{+-+0}(\s,\t,\u) = F_{+--0}(\s,\u,\t) \ . \nonumber
\eqa \par

There are three different forms of contributions to these
amplitudes arising from a $W$, fermion or scalar particle loop.
To express them economically, we use the notation of \cite{Hagiwara}
for the $B_0$, $C_0$ and $D_0$
1-loop functions first defined by Passarino and Veltman
\cite{Passarino}, and we introduce the shorthand writing
\bqa
B_0(\s)& \equiv  & B_0(\s;m,m) \ , \label{B0} \\
C_0(\s) & \equiv & C_0(12)=C_0(0,0,\s;m,m,m) \ , \label{C0}
\eqa
and
\bqa
B_Z(\s)& \equiv  & B_0(\s)-B_0(m^2_Z +i\epsilon) \ , \label{BZ} \\
C_Z(\s) & \equiv & C_Z(34)=C_0(m^2_Z,0,\s;m,m,m)
=C_0(0, m^2_Z,\s;m,m,m) \ , \label{CZ} \\
D_Z(\s,\u) & \equiv & D_Z(123)=D_0(0,0,0,m^2_Z,\s,\u;m,m,m,m)=D_Z(\u,\s)
\  \label{DZ}.
\eqa
The expressions
\bqa
\tilde{F}(\s,\t,\u)& \equiv  & D_Z(\s,\t)+D_Z(\s,\u)+D_Z(\u,\t)
\ \ , \label{Ftilde} \\
E(\t,\u)=E(\u, \t)&\equiv &\t C_0(\t)+\u C_0(\u)+\t_1 C_Z(\t)+\u_1
C_Z(\u)-\t\u D_Z(\t,\u) ,
\label{E}
\eqa
 appear naturally in the amplitudes below,
where  $\s_1=\s-m^2_Z$, $\t_1=\t-m^2_Z$, $\u_1=\u-m^2_Z$.\par

The $W$ loop contribution to the helicity amplitudes may then
be written as\footnote{The easiest way
to calculate this, is by using a non-linear gauge as
in \cite{Dicus}, in which the couplings $\gamma W^\pm\phi^\mp$,
$Z W^\pm\phi^\mp$ vanish. In such a gauge, the same propagator
appears along the entire loop.}  \cite{JikiaZ, Morgan}
\bq
F^W_{\lambda_1 \lambda_2 \lambda_3\lambda_4}(\s,\t,\u)\equiv
\alpha^2 ~{c_W\over s_W}
A^W_{\lambda_1 \lambda_2 \lambda_3\lambda_4} (\s,\t,\u)
\label{FWamp}
\eq
where
\bqa
&& A^W_{++++} (\s,\t,\u)= {16 \s_1\over \s} E(\t,\u)
+4[2(\s-4m^2_W)\s_1-m^2_W(m^2_Z-6m^2_W)]\tilde F(\s,\t,\u)
\nonumber \\
&&+2\left ({m^2_Z\over m^2_W}-6\right )
\Bigg \{{\t\u+m^2_W(\s+\s_1)\over \s\s_1}E(\t,\u)
-{2m^2_W\over \s_1}[\t\u D_Z(\t,\u)+m^2_ZC_0(\s)]\nonumber \\
&&-~{(\s+m^2_Z)\t\u\over \s_1\t_1\u_1}-{2m^2_Wm^2_Z\s\over
\s_1\t_1}C_Z(\t)
+\left({2\t+\s\over \s_1}-{m^4_Z\s\over \s_1\t^2_1}\right)
B_Z(\t)\nonumber \\
&&-~{2m^2_Wm^2_Z\s\over \s_1\u_1}C_Z(\u)
+\left ({2\u+\s\over \s_1}-{m^4_Z\s\over\s_1\u^2_1}\right )
B_Z(\u)\Bigg \} \ ,
\label{W++++}
\eqa
\bqa
&& A^W_{+++-} (\s,\t,\u)=2\left ({m^2_Z\over m^2_W}-6\right)
\Bigg \{-2m^4_W \tilde  F(\s,\t,\u)
-~{m^2_Z\t\u\over \s^2\s_1} E(\t,\u)\nonumber \\
&&+ ~ m^2_W \left [{(4m^2_Z-\s)\t\u\over
\s\s_1}D_Z(\t,\u)-{\s(\u^2+\t^2)\over \s_1\t\u}C_0(\s)-
{\s^2_1\over \u\t}C_Z(\s)\right ]
+ ~{(\s+m^2_Z)\t\u\over \s_1\t_1\u_1} \nonumber \\
&& +~ m^2_W \left [
\left({(m^2_Z\u-\s\t)\s\over \s_1\t_1\u}+{2m^2_Z\u-\s\u_1\over
\s_1\s}\right )C_Z(\t)-{(2m^2_Z\u+\s\t)\t\over
\s\u\s_1}C_0(\t) -{\s\t\over \u}D_Z(\s,\t) \right ]
\nonumber \\
&&+m^2_W \left [\left({(m^2_Z\t-\s\u)\s\over \s_1\u_1\t}
+{2m^2_Z\t-\s\t_1\over
\s_1\s}\right )C_Z(\u)-{(2m^2_Z\t+\s\u)\u\over
\s\t\s_1}C_0(\u)-{\s\u\over\t}D_Z(\s,\u)\right ] \nonumber \\
&& +{m^2_Z(2\t_1-\s)\u\t\over \s\s_1\t^2_1}B_Z(\t)
+{m^2_Z(2\u_1-\s)\u\t\over \s\s_1\u^2_1}B_Z(\u) \Bigg \}
 , \label{W+++-}
\eqa
\bqa
&& A^W_{++-+} (\s,\t,\u)=  2  \left ({m^2_Z\over
m^2_W}-6\right ) \Bigg \{-2 \mw^4 \tilde F(\s,\t,\u) +1
\nonumber \\
&& - \mwd \Big  [ {\t\u\over \s_1}D_Z(\t,\u)
 +{\s(\u^2+\t^2)\over \s_1\t\u}C_0(\s)
+{\s^2_1\over \u\t}C_Z(\s)
+ {\t\s\over \u}D_Z(\s,\t)
+{\u^2+\s^2_1 \over \s_1\u}C_0(\t)
\nonumber \\
&&+{\t\t_1\over \s_1\u}C_Z(\t)
+{\s\u\over \t}D_Z(\s,\u)+{\t^2+\s^2_1\over
\s_1\t}C_0(\u)+{\u\u_1\over \s_1\t}C_Z(\u)
\Big ] \Bigg \} ,
\label{W++-+}
\eqa
\bqa
 A^W_{++--} (\s,\t,\u) &= &
2\left ({m^2_Z\over m^2_W}-6 \right )
\{1-2m^4_W \tilde F(\s,\t,\u)
\nonumber \\
&& -{m^2_Wm^2_Z\over \s\s_1}[E(\u,\t)+2\s C_0(\s)]\} ,
\label{W++--}
\eqa
\bqa
&& A^W_{+-+-} (\s,\t,\u)= A^W_{+--+} (\s,\u,\t)=
 16 {\s\over \s_1} E(\s,\t)
\nonumber \\
&&+4\left({2\s\u(\u-4m^2_W)\over \s_1}-m^2_W(m^2_Z-6m^2_W)\right)
\tilde F(\s,\t,\u)\nonumber \\
&&+2\left ({m^2_Z\over m^2_W}-6\right)
\Bigg \{\left ({\s\t\over \u^2}+{2m^2_W\over \u}\right )E(\s,\t)
-m^2_W \Big [{2\s\t\over \u}D_Z(\s,\t)+{m^2_Z\over
\s\s_1}E(\u,\t)\nonumber \\
&&+{2m^2_Z(2\t_1+\s)\over \s_1\t_1}C_Z(\t)\Big ]+{\s(\s_1-\t)\over
\s_1\u}B_Z(\s)
-{\s\t(2\s_1\t_1+\t\u)\over \s_1\u\t^2_1}B_Z(\t)
-{\s\t\over \s_1\t_1} \Bigg \} ,
\label{W+-+-}
\eqa
\bqa
&& A^W_{+---} (\s,\t,\u)= A^W_{+-++} (\s,\u,\t)=
{8m^2_Z\u\over \s_1\t} [2E(\s,\u)
- \t(4m^2_W-\t)\tilde F(\s,\t,\u) ] \nonumber \\
&&-2 \left ({m^2_Z\over m^2_W}-6 \right)
\Bigg \{m^2_W \Big [\left ({\s\u\over \t}+2m^2_W \right)D_Z(\s,\u)
+\left ({\s\t\over\u}+2m^2_W \right )D_Z(\s,\t) \nonumber \\
&&+\left ({\t\u\over \s_1}+2m^2_W \right )D_Z(\t,\u)
+{\s\s_1\over \u\t}C_0(\s)
+{\u^2\over \s_1\t}C_0(\u)
+{\u^2+\s^2_1\over \s_1\u}C_0(\t)\nonumber \\
&&+{\u^2+\t^2\over \u\t}C_Z(\s)+{2m^2_Z\t^2
+\u_1(\t\t_1+\s\s_1)\over \s_1\u_1\t}C_Z(\u)
+{\t\t_1\over \s_1\u}C_Z(\t)\Big ]\nonumber \\
&&+{m^2_Z\t\u\over \s_1\u^2_1}B_Z(\u)
-{\s\u\over \s_1\u_1} \Bigg \} \ ,
\label{W+---}
\eqa
\bqa
&& A^W_{+++0} (\s,\t,\u)=p_t \sqrt{2}~ m_Z
\left ({m^2_Z\over m^2_W}-6 \right )
\Bigg \{(\t-\u)
\Big[{3m^2_W\over \s_1}D_Z(\t,\u)-{E(\t,\u)\over \s\s_1}
\nonumber \\
&& +{2m^2_W\s\over \s_1\t\u}C_0(\s)
 +{2\s\over \s_1\t_1\u_1} \Big ]
+m^2_W \left [{\s\over \u}D_Z(\s,\t)+{2\over \s_1}C_0(\t)
+{2(2\s^2-\t^2_1)\over \s_1\t_1\u}C_Z(\t)\right ]\nonumber \\
&&+{2(2\t_1\t+m^2_Z\u)\over \s_1\t^2_1}B_Z(\t)
-m^2_W \left [{\s\over \t}D_Z(\s,\u)+{2\over \s_1}C_0(\u)
+{2(2\s^2-\u^2_1)\over \s_1\u_1\t}C_Z(\u)\right ]\nonumber \\
&& -{2(2\u_1\u+m^2_Z\t)\over \s_1\u^2_1}B_Z(\u)\Bigg \} \ ,
\label{W+++0}
\eqa
\bqa
&& A^W_{++-0} (\s,\t,\u)=p_t \sqrt{2}~ m_Z (m^2_Z-6m^2_W)
\Big \{{(\t-\u)\over \s_1}
\left (D_Z(\t,\u)-{2\s\over \u\t}C_0(\s) \right )\nonumber \\
&&-{\s\over \u}D_Z(\s,\t)+{2\over \s_1}C_0(\t)
-{2\t_1\over \s_1\u}C_Z(\t)
+{\s\over \t}D_Z(\s,\u)-{2\over \s_1}C_0(\u)
+{2\u_1\over \s_1\t}C_Z(\u) \Big \} ,
\label{W++-0}
\eqa
\bqa
&& A^W_{+-+0} (\s,\t,\u)= A^W_{+--0} (\s,\u,\t)=
p_t \sqrt{2} \Bigg \{
{8 m_Z\s\over \s_1}
\left [(\u-4m^2_W)\tilde F(\s,\t,\u)
+{2E(\s,\t)\over \u}\right ] \nonumber \\
&& + m_Z \left ({m^2_Z\over m^2_W}-6 \right)
\Big  \{ m^2_W
\left [{(\t-\u)\over \s_1}D_Z(\t,\u)+{\s\over \t}D_Z(\s,\u)
+{3\s\over \u}D_Z(\s,\t)\right ]\nonumber \\
&&+{2m^2_W\over \s_1}
\left [C_0(\t)+C_0(\u)-{\s\s_1\over \u\t}C_0(\s)
+{\u_1\over \t}C_Z(\u)+{\t_1\over \u}C_Z(\t)
-{2\u\over \t_1}C_Z(\t)\right ] \nonumber \\
&& -{\s\over u^2}E(\s,\t)
 +{2\s\over \s_1\u}B_Z(\s)
-2 \left ({1\over \u}+{m^2_Z\u\over \s_1 \t^2_1} \right )B_Z(\t)
+{2\s\over \s_1\t_1} \Big \} \Bigg \} ,
\label{W+-+0}
\eqa
\noindent
where
\bq
p_t=\sqrt{{\t\u\over\s}} \ \ .
\label{pt}
\eq
When comparing these results with those of \cite{JikiaZ},
where helicity amplitudes are also given, we
identify some discrepancies in (\ref{W+++-},
\ref{W++-+}, \ref{W+++0}, \ref{W++-0}, \ref{W+-+0}), which are
of minor importance though, since the affected
amplitudes are very small. In addition there is a difference
in sign for the longitudinal $Z$ amplitudes, since \cite{JikiaZ}
does not use the Jakob-Wick convention\footnote{Note also that
the definitions of $\t$ and $\u$ used here and in \cite{gggg, GPRgamma1}
should be interchanged when comparing with \cite{JikiaZ}.} \par

We next turn to  the  \underline{fermion} loop contribution.
Writing the effective $Zf\bar f$ interaction as
\bq
\L_{Zff} = -e Z^\mu \bar f
(\gamma_\mu g_{Vf}^Z- \gamma_\mu \gamma_5 g_{Af}^Z) f ,
\label{LZff}
\eq
we remark that, due to charge conjugation,
only the vector coupling $g_{Vf}^Z$ gives a non-vanishing
contribution. For ordinary quarks and leptons this is
\bq
g_{Vf}^Z=\frac{t_3^f-2Q_f\sw^2}{2\sw\cw} \ ,
\label{gVfZ}
\eq
where $t_3^f$ is the fermion third isospin component, and
$Q_f$ its charge. Denoting then the fermion mass as $m_f$,
its contribution to the helicity
amplitudes is written as \cite{GloverZ}
\bq
F^f_{\lambda_1 \lambda_2 \lambda_3\lambda_4}(\s,\t,\u)\equiv
\alpha^2 Q^3_f g^Z_{Vf}
A^f_{\lambda_1 \lambda_2 \lambda_3\lambda_4} (\s,\t,\u) \ .
\label{Ffamp}
\eq
For presenting the fermion loop contribution it is convenient
to introduce the definitions
\bq
x_f\equiv {4m^2_f\over m^2_Z-6m^2_f} \ ,
\label{xf}
\eq
 and\footnote{The functions $E(\t,\u)$ and $\tilde F(\s,\t,\u)$
are defined  in (\ref{E}) and  (\ref{Ftilde}) respectively,
with $m=m_f$.}
\bq
G_f(\s,\t,\u)={(10m^2_f+m^2_Z)\over
m^2_f}E(\t,\u)+2\s[4\s-10m^2_f-m^2_Z]
\tilde{F}(\s,\t,\u) \ , \label{Gf}
\eq
which allow us to write \cite{GloverZ}
\bq
 A^f_{++++} (\s,\t,\u)=x_fA^W_{++++}(\s,\t,\u;~ m_W\to m_f)
- {x_f\s_1\over \s}G_f(\s,\t,\u)
 \  , \label{f++++}
\eq
\bqa
 A^f_{+++-} (\s,\t,\u)& = &x_fA^W_{+++-}(\s,\t,\u;~ m_W\to m_f)
\ , \label{f+++-} \\
 A^f_{++-+} (\s,\t,\u)& = & x_fA^W_{++-+}(\s,\t,\u;~ m_W\to m_f)
\ ,  \label{f++-+} \\
 A^f_{++--} (\s,\t,\u)& = &x_fA^W_{++--}(\s,\t,\u;~ m_W\to m_f)
\ , \label{f++--}
\eqa
\bqa
 A^f_{+-+-} (\s,\t,\u)=  A^f_{+--+} (\s,\u,\t)&= &
x_fA^W_{+-+-}(\s,\t,\u;  m_W\to m_f) \nonumber \\
&& -{x_f \s\over \s_1}G_f(\u,\t,\s)
   \ , \label{f+-+-} \\
  A^f_{+---} (\s,\t,\u)=  A^f_{+-++} (\s,\u,\t)& = &
x_fA^W_{+---}(\s,\t,\u; ~ m_W\to m_f) \nonumber \\
&& -{x_f \u m^2_Z\over \s_1\t}G_f(\t,\s,\u)
   , \label{f+---}
\eqa
\bqa
 A^f_{+++0} (\s,\t,\u)& = & x_f A^W_{+++0}(\s,\t,\u;~ m_W\to m_f)
\ , \label{f+++0} \\
 A^f_{++-0} (\s,\t,\u)& = & x_fA^W_{++-0}(\s,\t,\u; ~m_W\to m_f)
\ , \label{f++-0}
\eqa
\bqa
 A^f_{+-+0} (\s,\t,\u)= A^f_{+--0} (\s,\u,\t) &= &
x_f A^W_{+-+0}(\s,\t,\u;~ m_W\to m_f) \nonumber \\
&& -{x_f m_Z\over \s_1}\sqrt{{2\t\s\over \u}}G_f(\u,\t,\s)
 \  . \label{f+-+0}
\eqa \par

 Finally the contribution to the helicity amplitudes arising from
a loop due to  a \underline{scalar} particle\footnote{Like \eg\@ a
slepton.} of  charge $Q_{S}$, mass $m_S$, and third isospin component
$t^S_3$,  is
\bq
F^S_{\lambda_1 \lambda_2 \lambda_3\lambda_4}(\s,\t,\u)\equiv
\alpha^2 Q^3_S g^Z_{S}
A^S_{\lambda_1 \lambda_2 \lambda_3\lambda_4} (\s,\t,\u) \ ,
\eq
where
\bq
g^Z_{S}  =  \frac{t^S_3-Q_S \sw^2}{\sw\cw}  \ \ . \label{gSZ}
\eq
Using then the definitions
\bqa
x_S &=&  {2m^2_S\over 6m^2_S-m^2_Z} \ \ , \label{xS} \\
G_S(\s,\t,\u)& = & 2E(\t,\u)+\s(\s-4m^2_S)\tilde{F}(\s,\t,\u)
\ , \label{GS}
\eqa
we obtain
\bq
 A^S_{++++} (\s,\t,\u)=x_S A^W_{++++}(\s,\t,\u;~ m_W\to m_S)
- {8 x_S\s_1\over \s}G_S(\s,\t,\u)
 \  , \label{S++++}
\eq
\bqa
 A^S_{+++-} (\s,\t,\u)& = &x_S A^W_{+++-}(\s,\t,\u;~ m_W\to m_S)
\ , \label{S+++-} \\
 A^S_{++-+} (\s,\t,\u)& = & x_S A^W_{++-+}(\s,\t,\u;~ m_W\to m_S)
\ ,  \label{S++-+} \\
 A^S_{++--} (\s,\t,\u)& = &x_S A^W_{++--}(\s,\t,\u;~ m_W\to m_S)
\ , \label{S++--}
\eqa
\bqa
 A^S_{+-+-}(\s,\t,\u)=  A^S_{+--+} (\s,\u,\t)&= &
x_S A^W_{+-+-}(\s,\t,\u;  m_W\to m_S) \nonumber \\
&& -{8 x_S \s\over \s_1}G_S(\u,\t,\s)
   \ , \label{S+-+-} \\
A^S_{+---} (\s,\t,\u)=  A^S_{+-++} (\s,\u,\t)& = &
x_S A^W_{+---}(\s,\t,\u; ~ m_W\to m_S) \nonumber \\
&& -{8 x_S \u m^2_Z\over \s_1\t}G_S(\t,\s,\u)
   , \label{S+---}
\eqa
\bqa
 A^S_{+++0} (\s,\t,\u)& = & x_S A^W_{+++0}(\s,\t,\u;~ m_W\to m_S)
\ , \label{S+++0} \\
 A^S_{++-0} (\s,\t,\u)& = & x_S A^W_{++-0}(\s,\t,\u; ~m_W\to m_S)
\ , \label{S++-0}
\eqa
\bqa
 A^S_{+-+0} (\s,\t,\u)= A^S_{+--0} (\s,\u,\t) &= &
x_S A^W_{+-+0}(\s,\t,\u;~ m_W\to m_S) \nonumber \\
&& -{8 x_S m_Z\over \s_1}\sqrt{{2\t\s\over \u}}G_S(\u,\t,\s)
 \  . \label{S+-+0}
\eqa \par

\vspace{2cm}


\renewcommand{\theequation}{B.\arabic{equation}}
\renewcommand{\thesection}{B.\arabic{section}}
\setcounter{equation}{0}
\setcounter{section}{0}

{\large \bf Appendix B: The asymptotic
$\gamma \gamma \to \gamma Z $ amplitudes in SM.}

At high energies the 1-loop functions simplify considerably.
Such asymptotic expressions are very useful in elucidating the
physical properties of the amplitudes at high energies; as can be
seen from \cite{gggg} for the
$\gamma \gamma \to \gamma \gamma $ case.
In this Appendix, we present therefore the
asymptotic expression for 1-loop functions relevant for the
$\gamma \gamma \to \gamma Z$ amplitudes.\par

Using thus the well known asymptotic
expression for the $B_0$ function of (\ref{B0})
\bq
B_0(\s) \simeq  \Delta +2 -\ln \left (\frac{-\s -i\epsilon}{\mu^2} \right )
\ ~ , \label{B0asym}
\eq
where $\Delta $ is the usual infinite term entering the calculation
of  the  divergent integral \cite{Hagiwara}, we obtain
for the $B_Z(\s)$ function defined in (\ref{BZ})
\bq
B_Z(\s) \simeq -\ln \left ( \frac{-\s-i \epsilon}
{-\mzd -i \epsilon}\right ) \ , ~~ \mbox{ for} ~~~~
~ |\s| \gg (m^2,~\mzd)\ \ . \label{BZasym}
\eq
 For
the $C_0(\s)$ function defined
in (\ref{C0}), a useful form is  \cite{Denner}
\bq
C_0(\s)  =  \frac{1}{2 \s} \left [\ln \left (\frac{-\s -i\epsilon}{m^2}
\right ) \right ]^2 \ \ . \label{C0asym}
\eq\par

We next turn to $C_Z(\s)$ and $D_Z(\s,\u)$ of(\ref{CZ}, \ref{DZ}),
which also depend on $m/\mz$.
Simple asymptotic expression are derived for them,
 for arbitrary $m/\mz$, using the results of  \cite{Denner}.
To present them we introduce the quantity
\bq
a_Z \equiv   \sqrt{ 1-~\frac{4 m^2}{\mzd} ~+i \epsilon}
\label{a-param}
\eq
Then, for $|\s| \gg (m^2,~\mzd)$,
\bqa
C_Z(\s)& \simeq & \frac{1}{\s} \Bigg \{
\frac{1}{2}\ln^2\left (\frac{-\s-i\epsilon}{m^2}\right)
+ \frac{\pi^2}{2}
-\frac{1}{2} \left[ \ln \left(\frac{1+a_Z}{2}\right )
- \ln \left(\frac{1-a_Z}{2}\right ) \right ]^2
\nonumber \\
&& + i\pi \left[ \ln \left(\frac{1+a_Z}{2}\right )
- \ln \left(\frac{1-a_Z}{2}\right )  \right ]
+ O\left (\frac{1}{\s} \right )  \Bigg \} \ \ ,
\label{CZasym}
\eqa
while for  $(|\s|,~ |\u|) \gg (m^2,~\mzd)$,
\bqa
D_Z(\s, \u)& \simeq & \frac{2}{\s\u} \Bigg \{
\ln \left (\frac{-\s-i\epsilon}{m^2}\right)
\ln \left (\frac{-\u-i\epsilon}{m^2}\right)
-\frac{1}{2} \left[ \ln \left(\frac{1+a_Z}{2}\right )
- \ln \left(\frac{1-a_Z}{2}\right ) \right ]^2
\nonumber \\
&& + i\pi \left[ \ln \left(\frac{1+a_Z}{2}\right )
- \ln \left(\frac{1-a_Z}{2}\right ) \right ]
+ O\left (\frac{1}{\s}~,~ \frac{1}{\u} \right ) \Bigg \} \ \ .
\label{DZasym}
\eqa
In all cases the principal value of the logarithm
is understood, which has its cut is along the
negative real axis.\par

We next turn to the functions $\tilde F(\s,\t,\u)$ and $E(\t, \u)$
defined in (\ref{Ftilde}, \ref{E}), which together with
$B_Z(\s)$ (compare (\ref{BZasym})),
determine the high energy behaviour of the various $\gamma \gamma
\to \gamma Z$ amplitudes. Using (\ref{CZasym}, \ref{DZasym}), they
can be expressed  for $(|\s|,~|\t|, ~|\u|) \gg (m^2, ~ \mzd)$ as
\bqa
\tilde F(\s, \t,\u) & \simeq &
\frac{2}{\s\u} \ln \left (\frac{-\s-i\epsilon}{m^2}\right)
\ln \left (\frac{-\u-i\epsilon}{m^2}\right)
+ \frac{2}{\s\t} \ln \left (\frac{-\s-i\epsilon}{m^2}\right)
\ln \left (\frac{-\t-i\epsilon}{m^2}\right)
\nonumber \\
&&
+\frac{2}{\t\u} \ln \left (\frac{-\t-i\epsilon}{m^2}\right)
\ln \left (\frac{-\u-i\epsilon}{m^2}\right) \ ,
\label{Ftildeasym} \\
E(\t,\u) & \simeq & \pi^2  + \left [
\ln\left ( \frac{-\t-i\epsilon}{m^2} \right )
-\ln\left ( \frac{-\u-i\epsilon}{m^2} \right ) \right ]^2
\ . \label{Easym}
\eqa
It is worth remarking  that no $a_Z$ term appears
in  (\ref{BZasym}), (\ref{Ftildeasym}) and
(\ref{Easym}).  This implies
that the asymptotic
$\gamma \gamma \to \gamma Z$ amplitudes do not depend on
the ratio $\mz/m$; as opposed to the situation
for the asymptotic  $C_Z$ and $D_Z$
functions.\par

The corresponding asymptotic expressions of the $W$
loop contributions obtained from (\ref{W++++}-\ref{W+-+0})
for $(|\s|,~|\t|, ~|\u|) \gg (\mwd,~  \mzd)$, by neglecting
 terms of $O(\mw^2/\s)$ but keeping terms of
$O(\mw/\sqrt{s})$ are
\bqa
&& A^W_{++++} (\s,\t,\u)  \simeq  16  E(\t,\u)
+ 8\s^2 \tilde F(\s,\t,\u)
\nonumber \\
&& +2\left ({1\over \cw^2}-6\right )
\left [{\t\u \over \s^2}E(\t,\u) -1+\frac{(\t-\u)}{\s}
[B_Z(\t)-B_Z(\u)] \right ]
\ , \label{W++++asym} \\
&& A^W_{+-+-} (\s,\t,\u)= A^W_{+--+} (\s,\u,\t) \simeq
 16  E(\s,\t) +8\u^2 \tilde F(\s,\t,\u) \nonumber \\
&&+2\left ({1\over \cw^2}-6\right)
 \left [{\t\s \over \u^2}E(\s,\t) -1+\frac{(\s-\t)}{\u}
[B_Z(\s)-B_Z(\t)] \right ]  ,
\label{W+-+-asym} \\
&& A^W_{+++0} (\s,\t,\u)=p_t \sqrt{2}~ m_Z
\left ({1\over \cw^2}-6 \right )
\Big [ (\t-\u)
\left ( -~\frac{E(\t,\u)}{\s^2} +\frac{2}{\t\u} \right )
\nonumber \\
&& +\frac{4}{\s} [B_Z(\t)-B_Z(\u)] \Big ]
 \ , \label{W+++0asym} \\
&& A^W_{+-+0} (\s,\t,\u)= A^W_{+--0} (\s,\u,\t)=
p_t \sqrt{2} ~\mz \Bigg \{ 8\u \tilde F(\s,\t,\u)
+\frac{16}{\u}E(\s,\t) \nonumber \\
&& + \left ({1\over \cw^2}-6 \right )
\left [-~\frac{\s}{\u^2} E(\s,\t)+\frac{2}{\u} [B_Z(\s)-B_Z(\t)]
+\frac{2}{\t} \right ]  \Bigg \} ,
\label{W+-+0asym}\\
&& A^W_{+++-}(\s, \t,\u)  \simeq A^W_{++-+}(\s, \t,\u) \simeq
 A^W_{++--}(\s, \t,\u)  \nonumber \\
&& \simeq  A^W_{+---}(\s, \t,\u) \simeq
 A^W_{+-++}(\s, \t,\u) \simeq 2 \left ({1\over \cw^2}-6 \right )
\ , \label{W+++-asym} \\
&& A^W_{++-0} \simeq 0 \ \ . \label{W++-0asym}
\eqa

It is easy to see that at energies above 250~GeV,
the Sudakov-like log-squared terms in
(\ref{Easym}, \ref{Ftildeasym}) largely cancel out, when substituted in
these asymptotic amplitudes. Essentially
only the single logarithm
large imaginary terms remain contributing to the dominant
amplitudes $A^W_{++++}(\s,\t,\u)$ and
$A^W_{+-+-}(\s,\t,\u)=A^W_{+--+}(\s,\u,\t)$.
Almost negligible are, the
the amplitudes in (\ref{W+++0asym}, \ref{W+-+0asym}), while the rest
are even smaller.

Similar asymptotic expression can also be obtained for
the fermion  loop contributions appearing
in (\ref{f++++} - \ref{f+-+0}), by taking
$(|\s|,~|\t|, ~|\u|) \gg (m_f^2,~  \mzd)$, and using
(\ref{Ftildeasym}, \ref{Easym}). It turns out that at energies above
250GeV, the fermion loop contribution in SM to the large imaginary
parts of the amplitudes $F_{++++}(\s,\t,\u)$ and
$F_{+-+-}(\s,\t,\u)=F_{+--+}(\s,\u,\t)$ is completely negligible.
Only to the other very small amplitudes, the W and fermion loop
contributions are comparable. \par

\begin{figure}[p]
\vspace*{-4cm}
\[
\epsfig{file=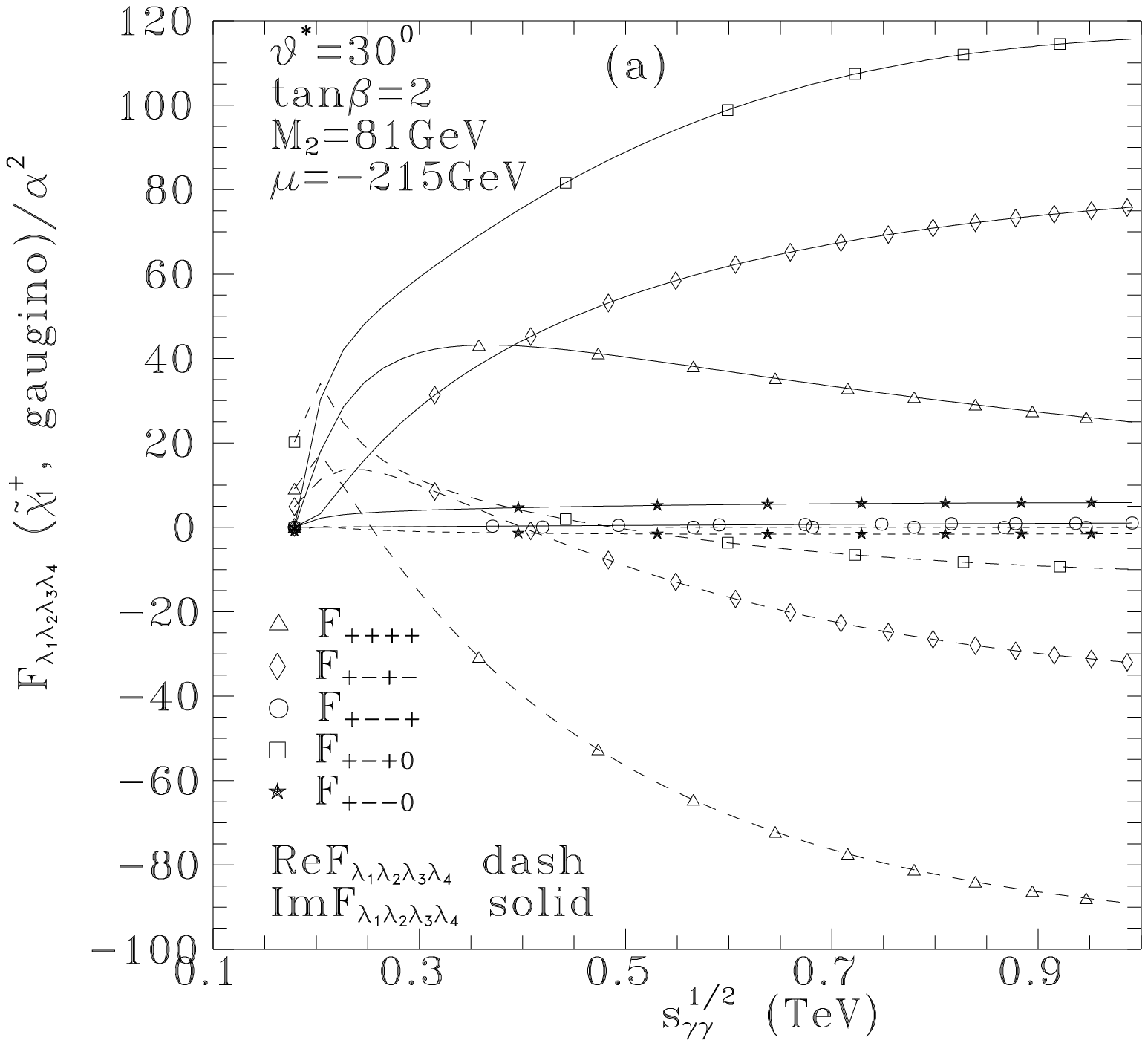,height=7.5cm}\hspace{0.5cm}
\epsfig{file=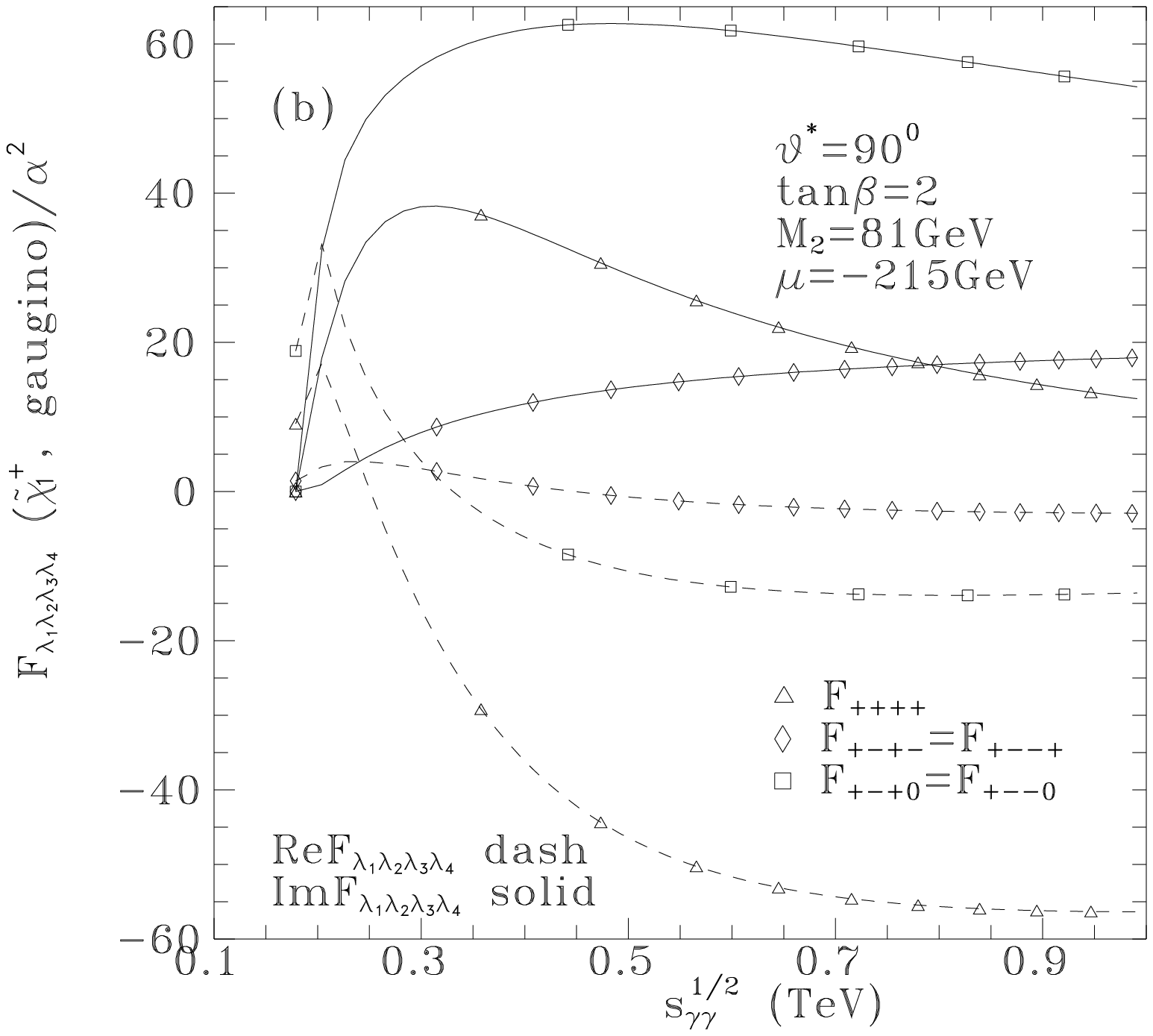,height=7.5cm}
\]
\vspace*{0.5cm}
\[
\epsfig{file=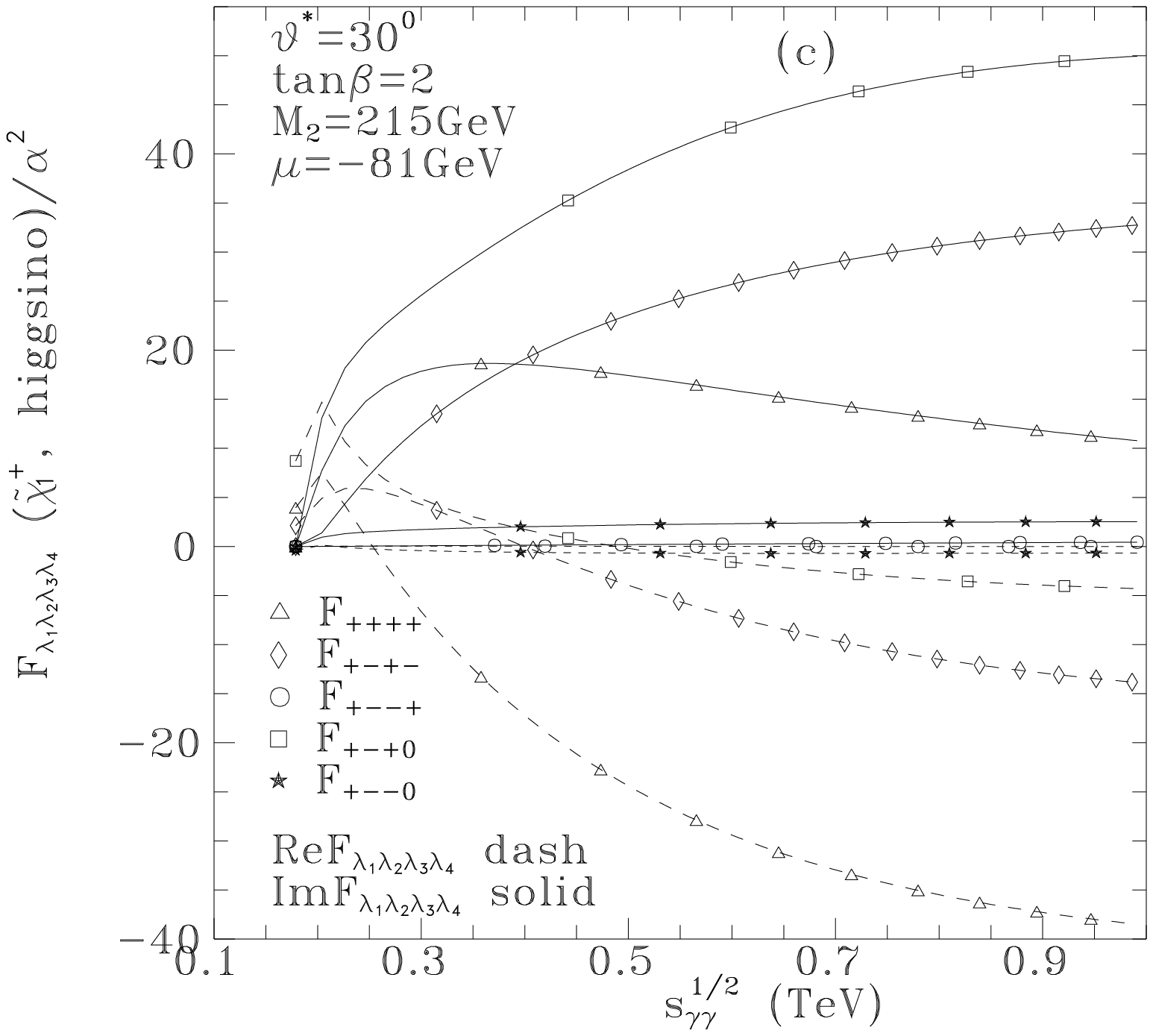,height=7.5cm}\hspace{0.5cm}
\epsfig{file=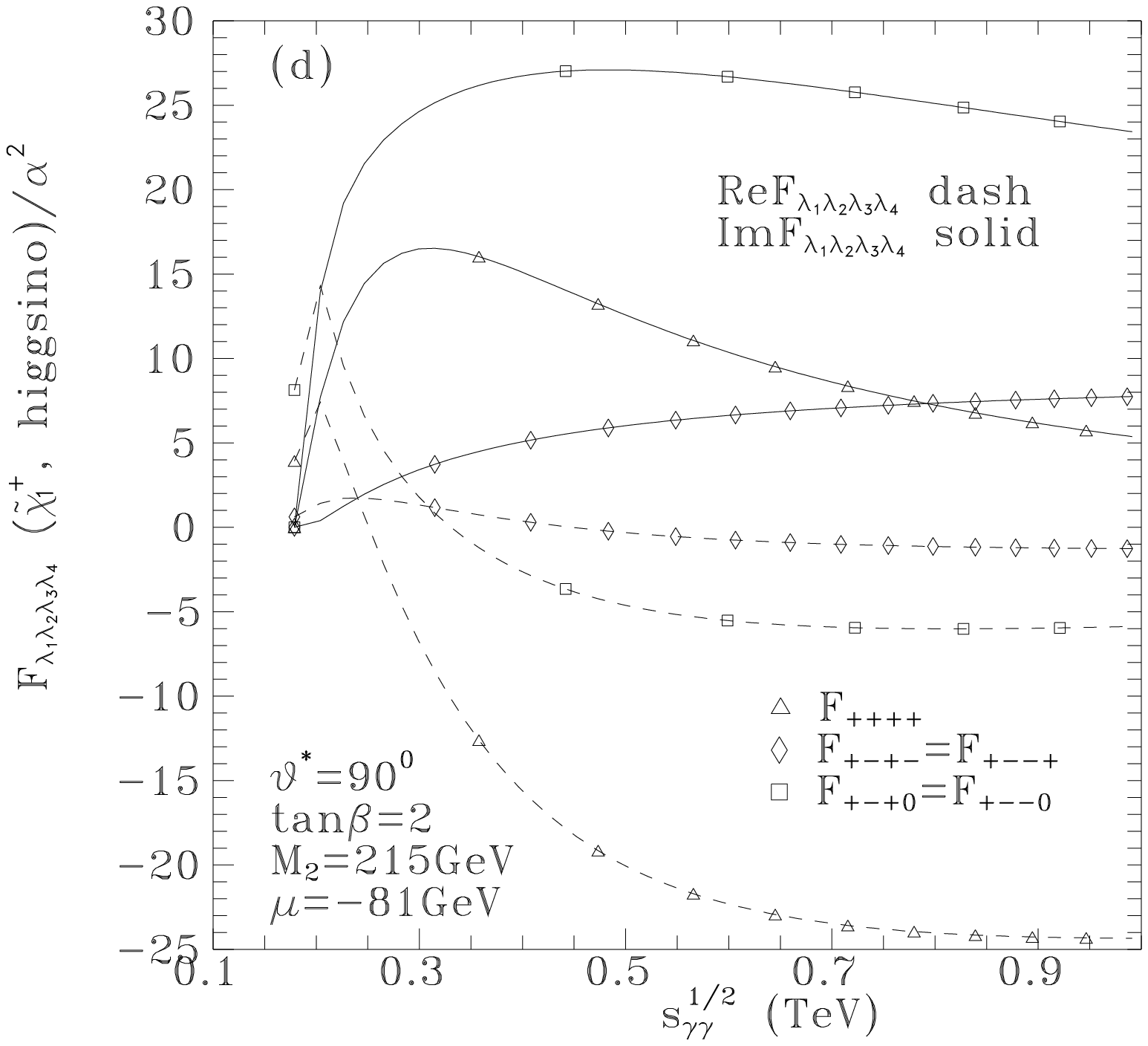,height=7.5cm}
\]
\vspace*{0.5cm}
\caption[1]{Chargino contribution to $\gamma \gamma \to \gamma Z$
helicity amplitudes for the gaugino  (a,b) and higgsino (c,d)
cases  at $\vartheta^*=30^0$ and $\vartheta^*=90^0$. The
parameters used are indicated in the figures
and $Q_{\chi^+_1}=1$.}
\label{chargino-amp}
\end{figure}

\begin{figure}[p]
\vspace*{-4cm}
\[
\epsfig{file=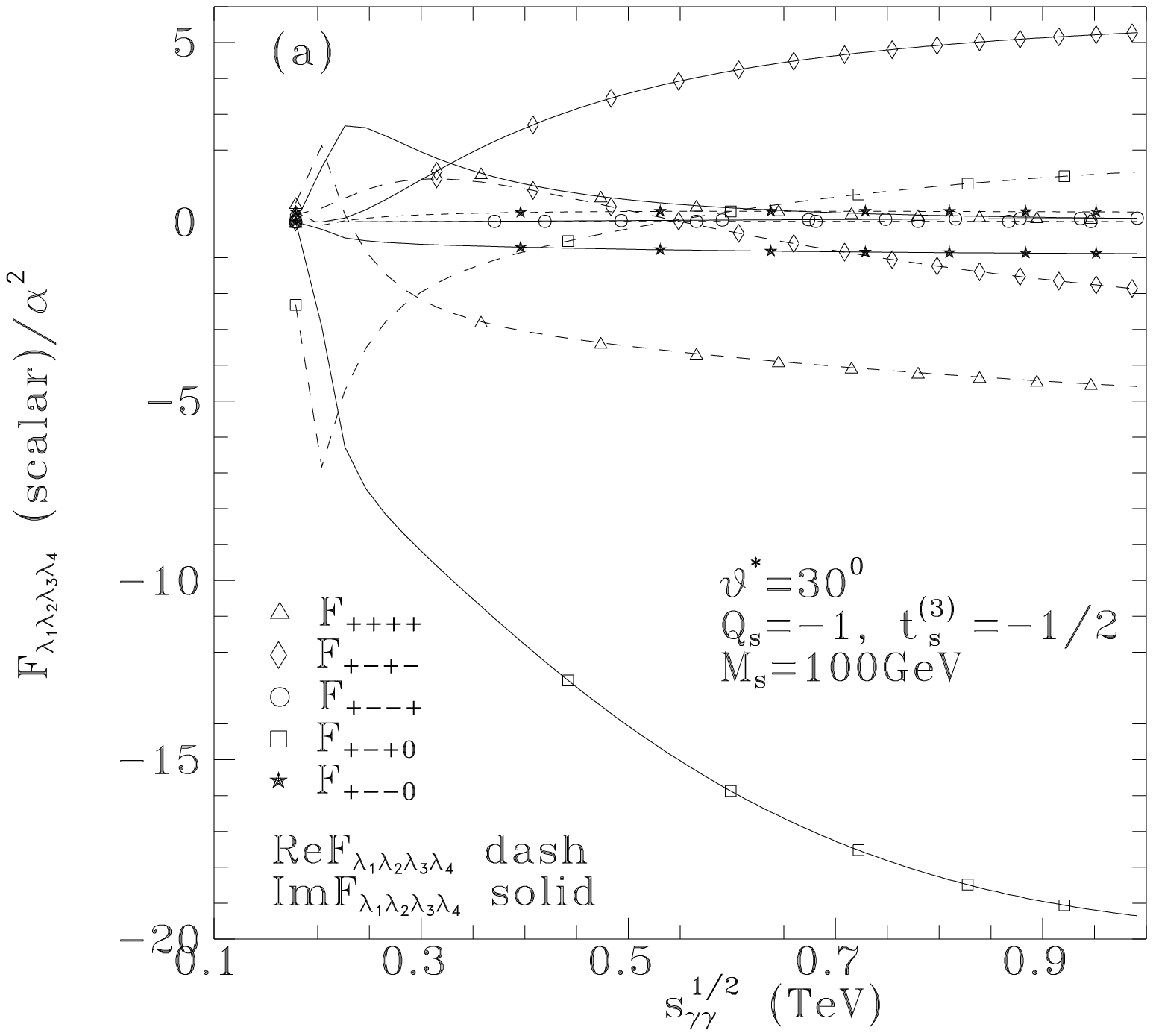,height=7.5cm}\hspace{0.5cm}
\epsfig{file=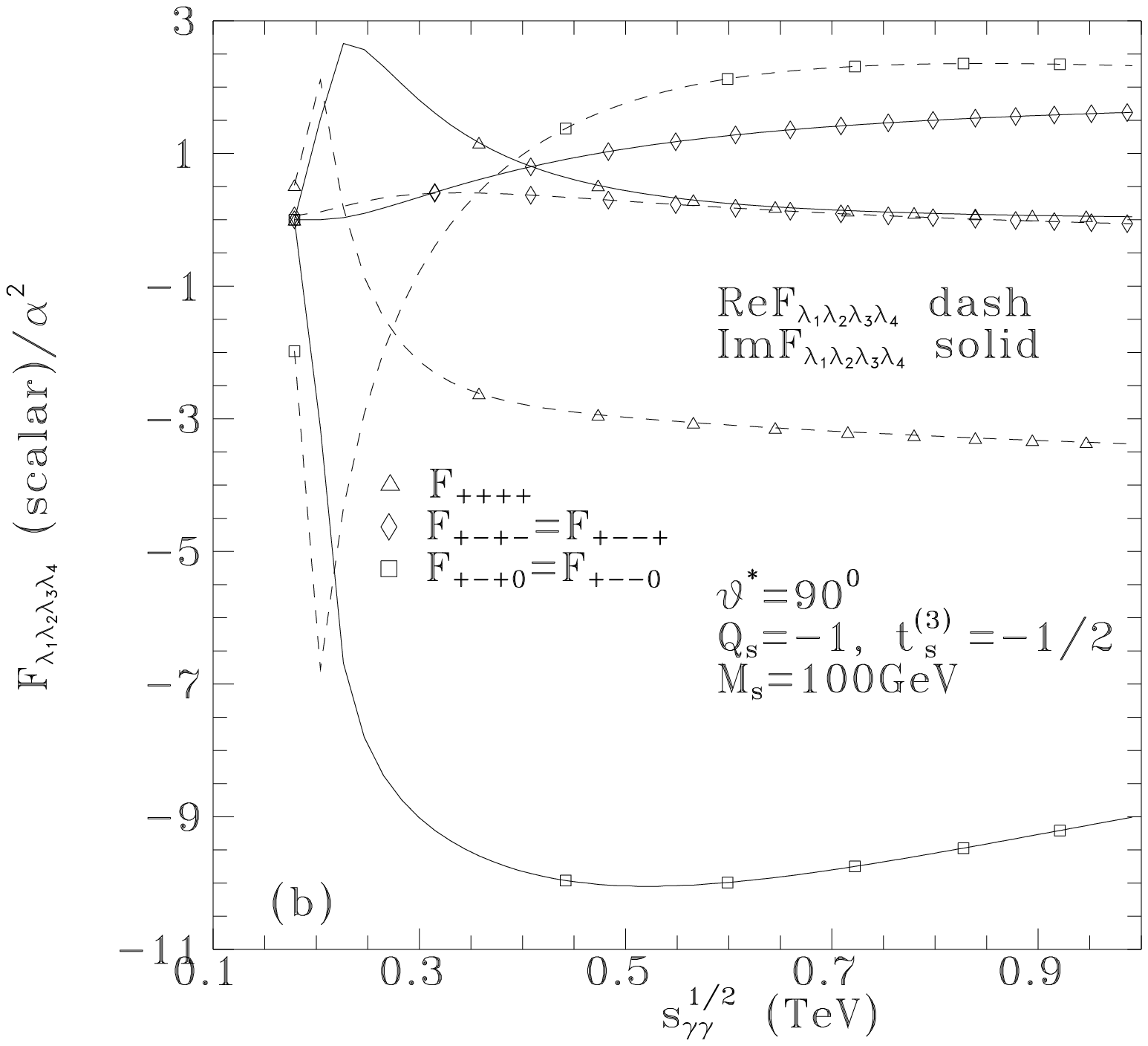,height=7.5cm}
\]
\vspace*{0.5cm}
\[
\epsfig{file=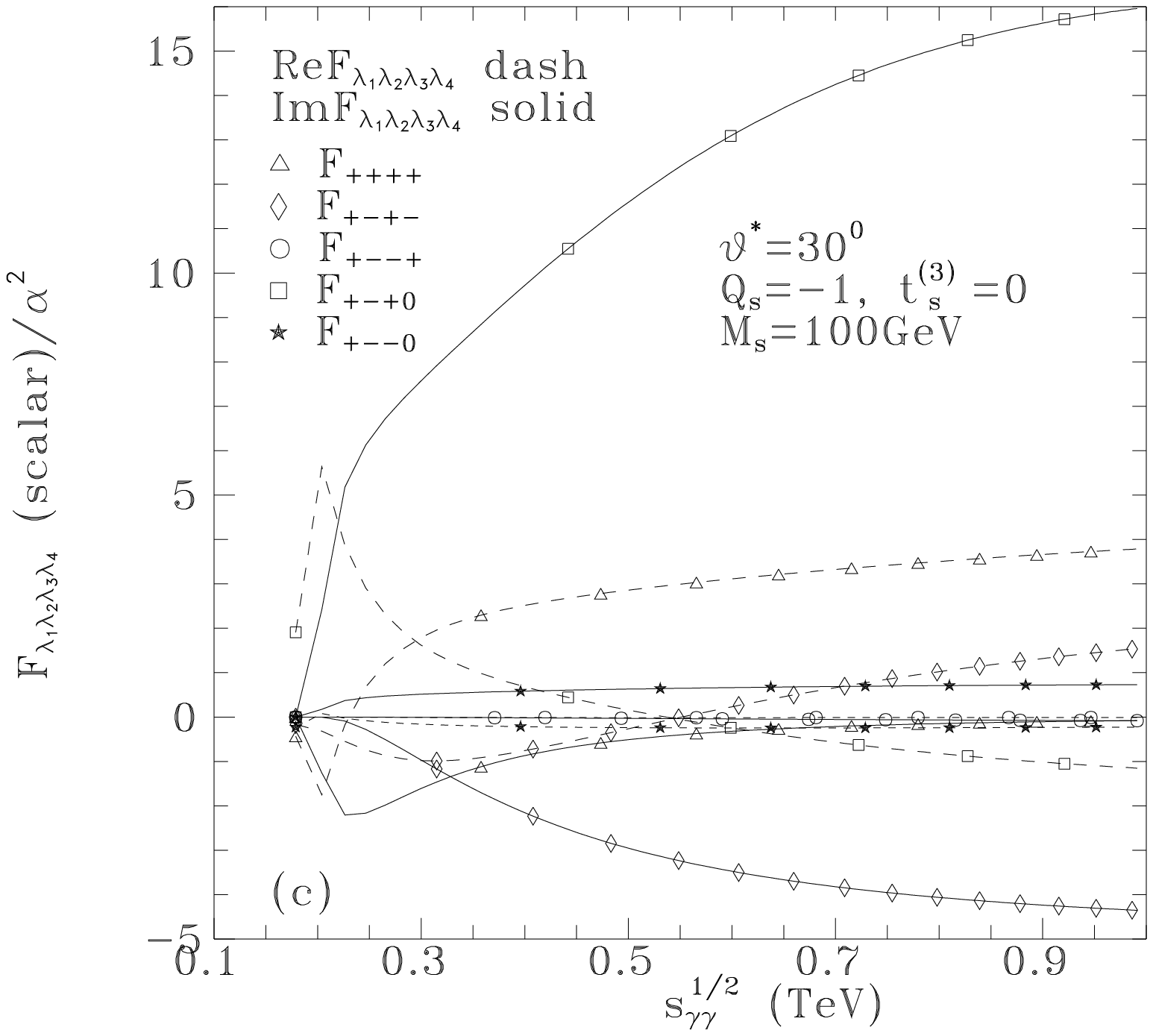,height=7.5cm}\hspace{0.5cm}
\epsfig{file=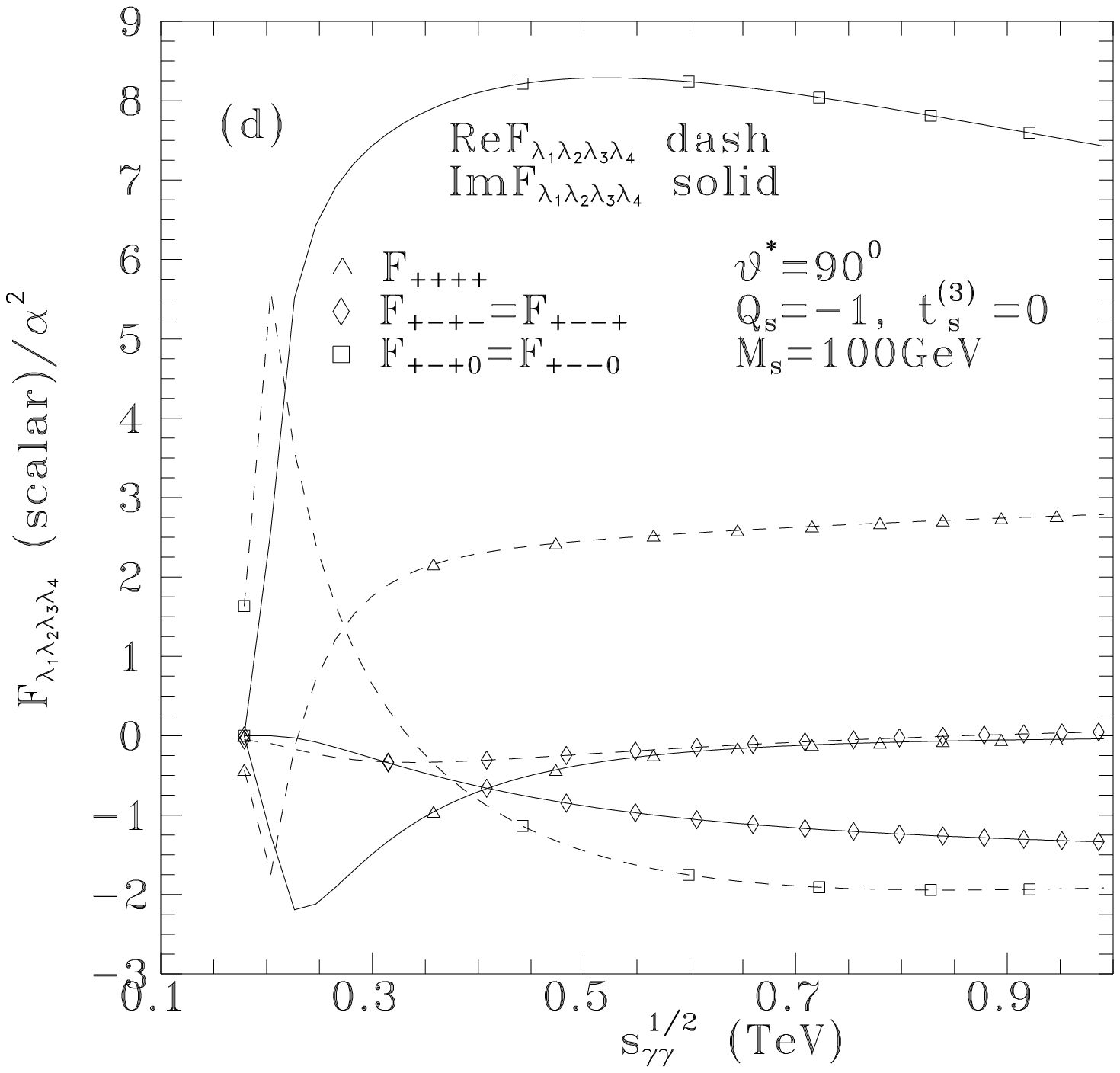,height=7.5cm}
\]
\vspace*{0.5cm}
\caption[1]{Contribution to $\gamma \gamma \to \gamma Z$
helicity amplitudes from an isodoublet (a,b) and an isosinglet
(c,d) slepton at $\vartheta^*=30^0$ and $\vartheta^*=90^0$.
The parameters used are indicated in the figures and the slepton mass
is taken $M_{\tilde l}=M_s=100GeV$. }
\label{slepton-amp}
\end{figure}

\begin{figure}[p]
\vspace*{-4cm}
\[
\epsfig{file=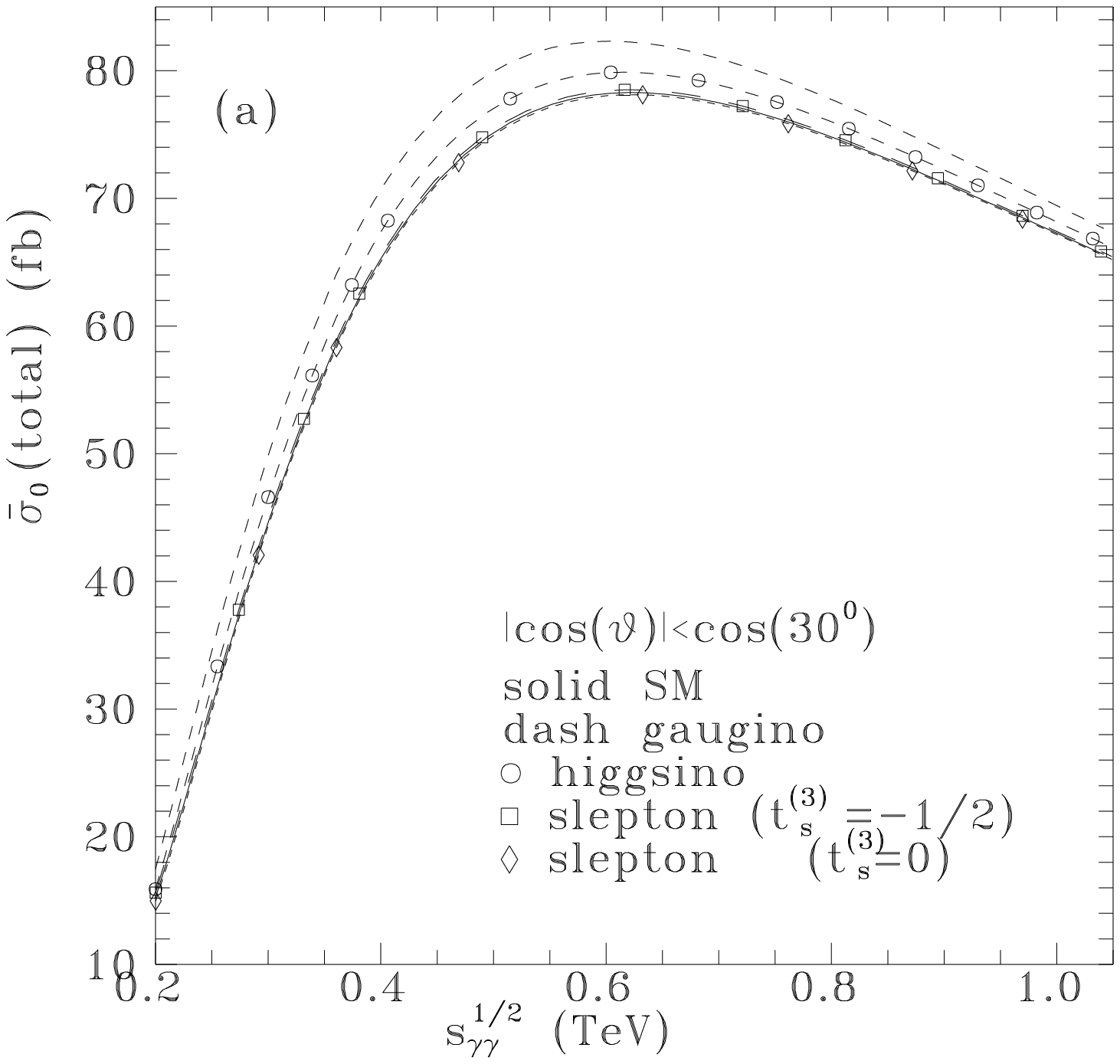,height=7.5cm}\hspace{0.5cm}
\epsfig{file=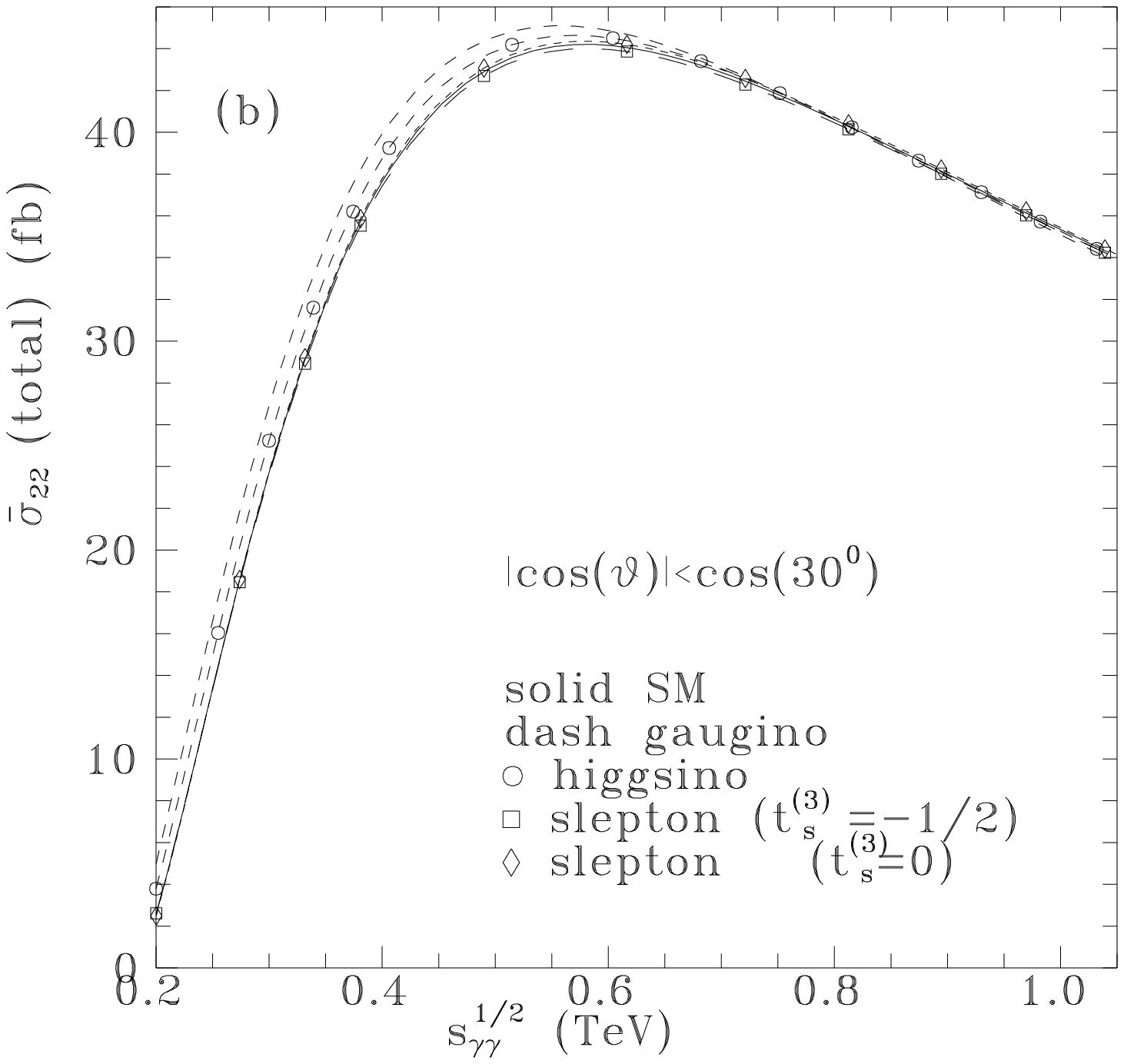,height=7.5cm}
\]
\vspace*{0.5cm}
\[
\epsfig{file=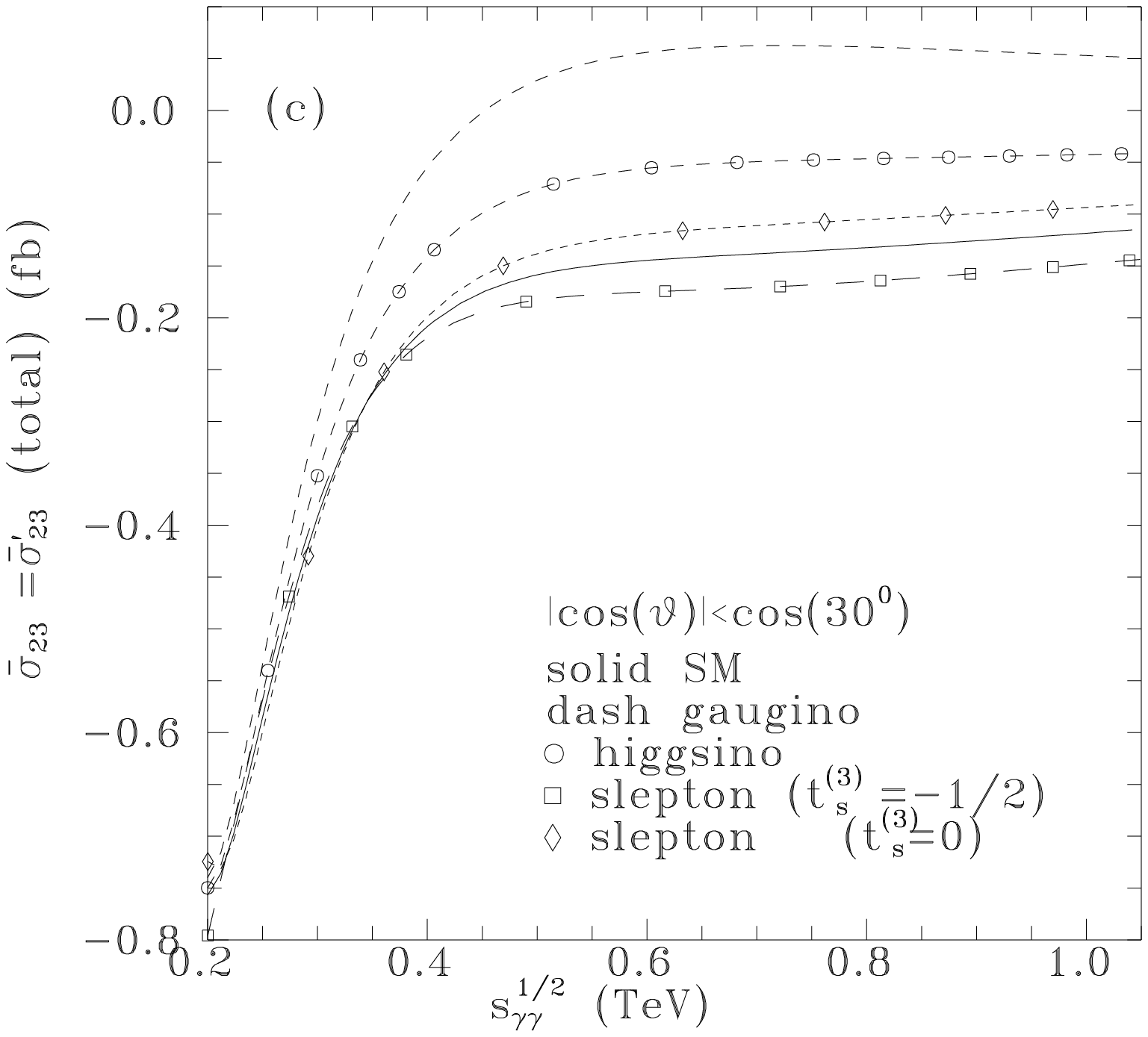,height=7.5cm}\hspace{0.5cm}
\epsfig{file=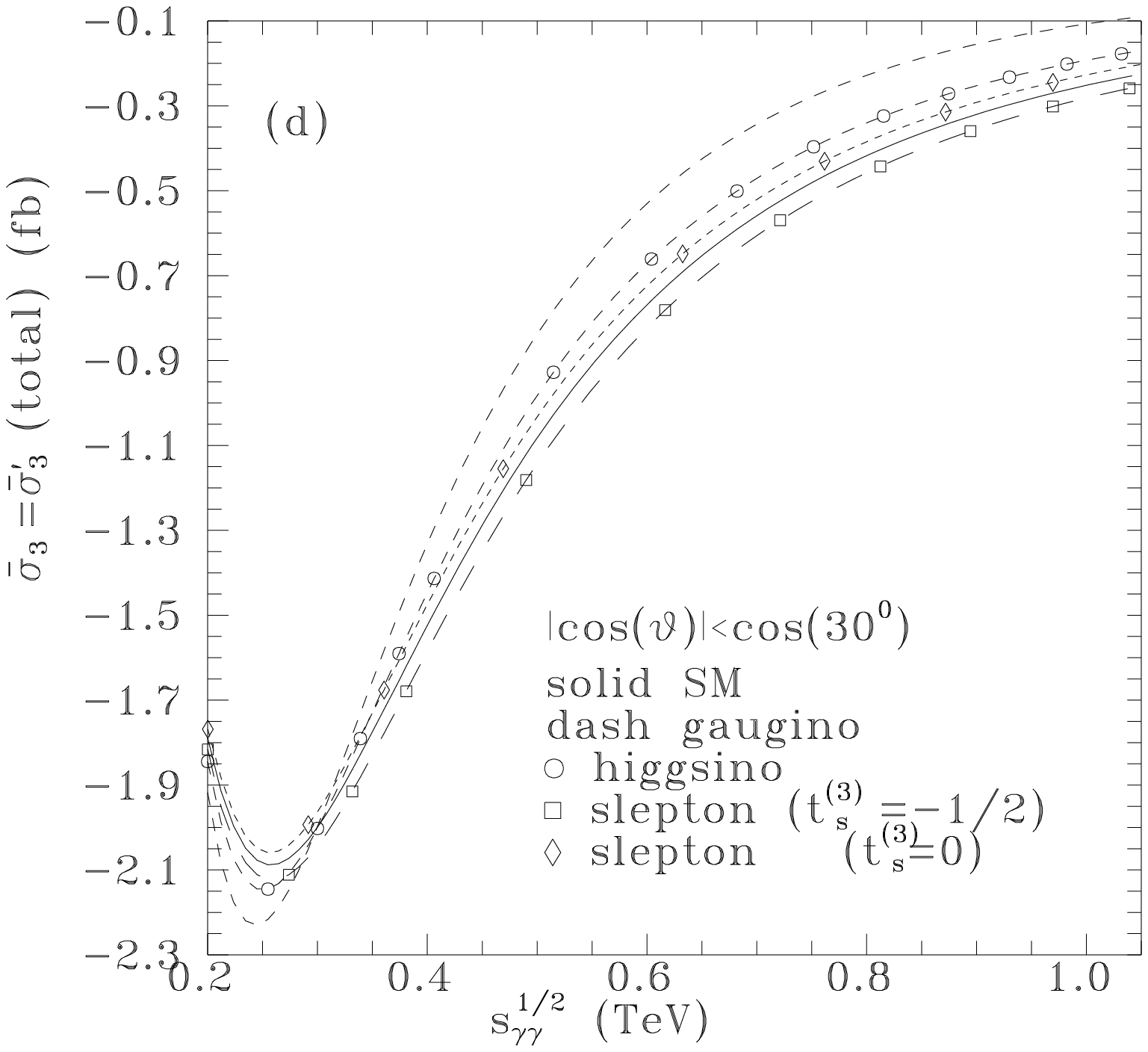,height=7.5cm}
\]
\vspace*{0.5cm}
\caption[1]{$\bar \sigma_0$, $\bar \sigma_{22}$,
$\bar \sigma_{23}=\bar \sigma_{23}^\prime $
and  $\bar \sigma_{3}=\bar\sigma_3^\prime $ for SM (solid) and
in the presence of a chargino (dash, dash-circle) or a charged slepton
(box, rhombus) contribution, using the same parameters as in
Fig.\ref{chargino-amp} or Fig.\ref{slepton-amp} respectively.}
\label{sig}
\end{figure}

>From Fig.\ref{chargino-amp} and
Fig.\ref{slepton-amp}, it can also be concluded that the real and
imaginary contributions of a fermion or scalar loop, are on an
equal footing. As said already, the large imaginary  contributions to
$F_{++++}(\s,\t,\u)$ and $F_{+-+-}(\s,\t,\u)=F_{+--+}(\s,\u,\t)$
comes from the dominant W loop only.

\clearpage
\newpage

\addtocounter{figure}{-1}

\begin{figure}[p]
\vspace*{-4cm}
\[
\epsfig{file=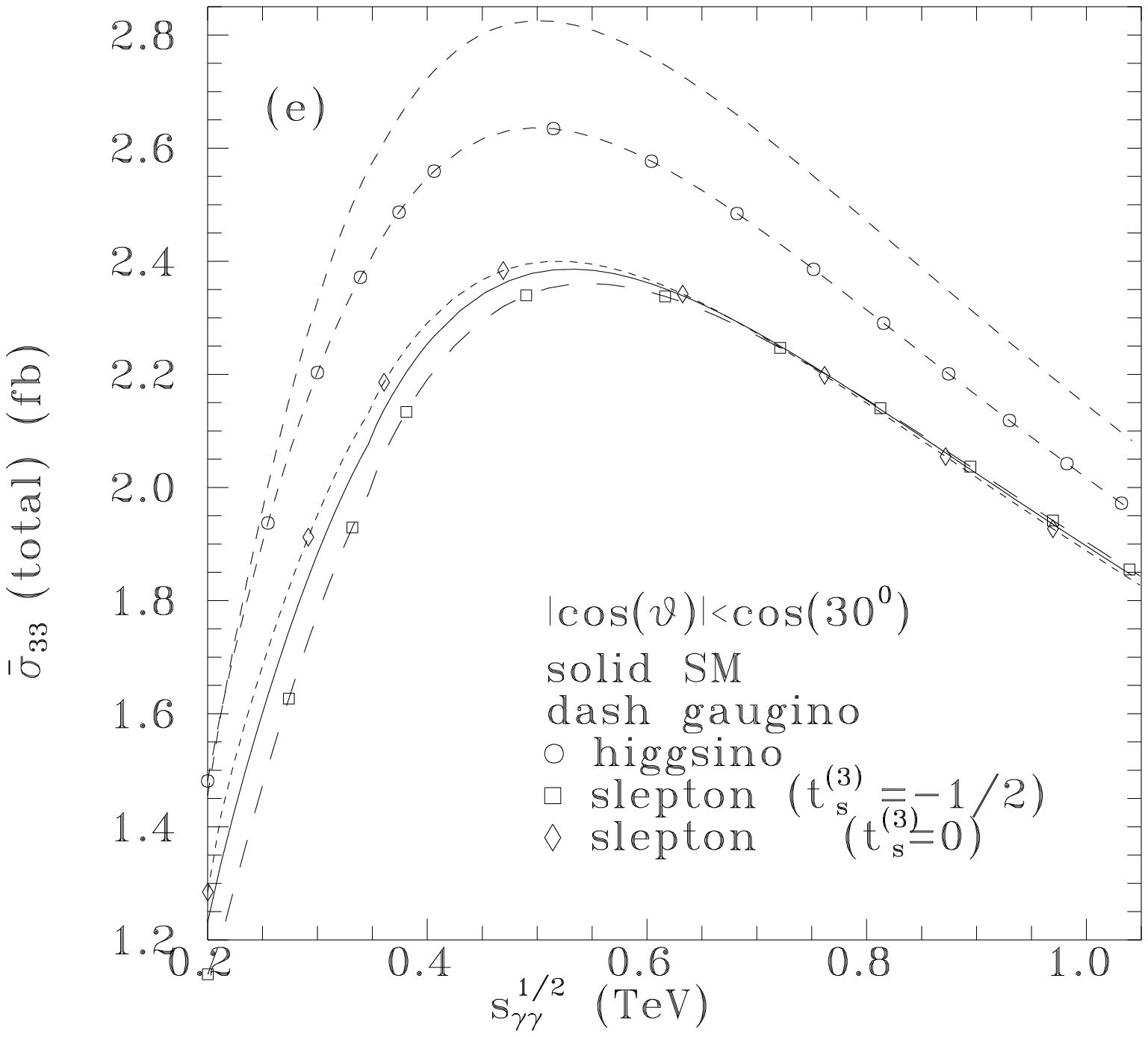,height=7.5cm}\hspace{0.5cm}
\epsfig{file=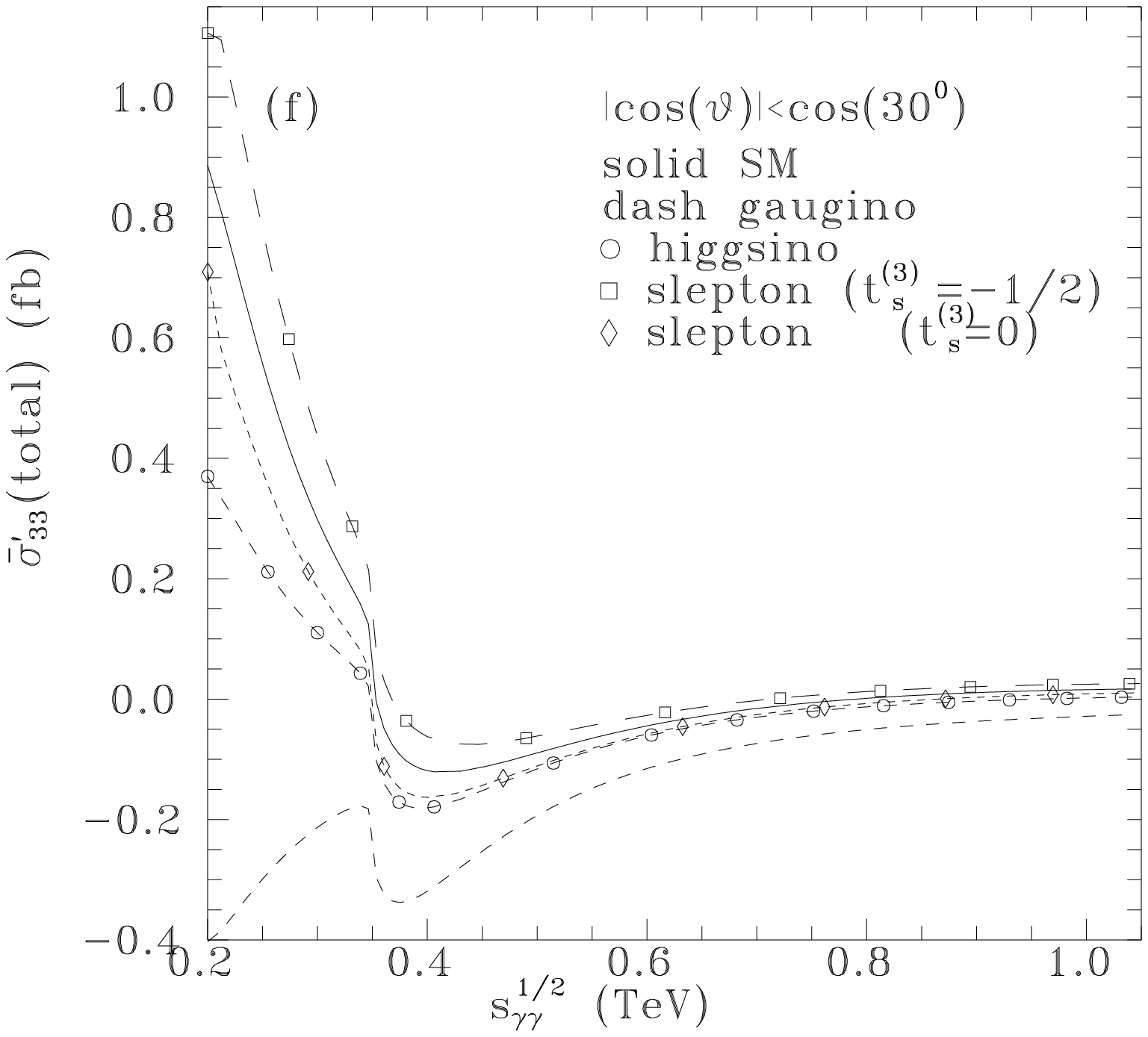,height=7.5cm}
\]
\caption[1]{$\bar \sigma_{33} $
and  $\bar \sigma'_{33} $ for SM (solid) and in the presence of
a chargino (dash, dash-circle) or a charged slepton
(box, rhombus) contribution, using the same parameters as in
Fig.\ref{chargino-amp} or Fig.\ref{slepton-amp} respectively.}
\label{sig1}
\end{figure}

\begin{figure}[p]
\vspace*{-4cm}
\[
\epsfig{file=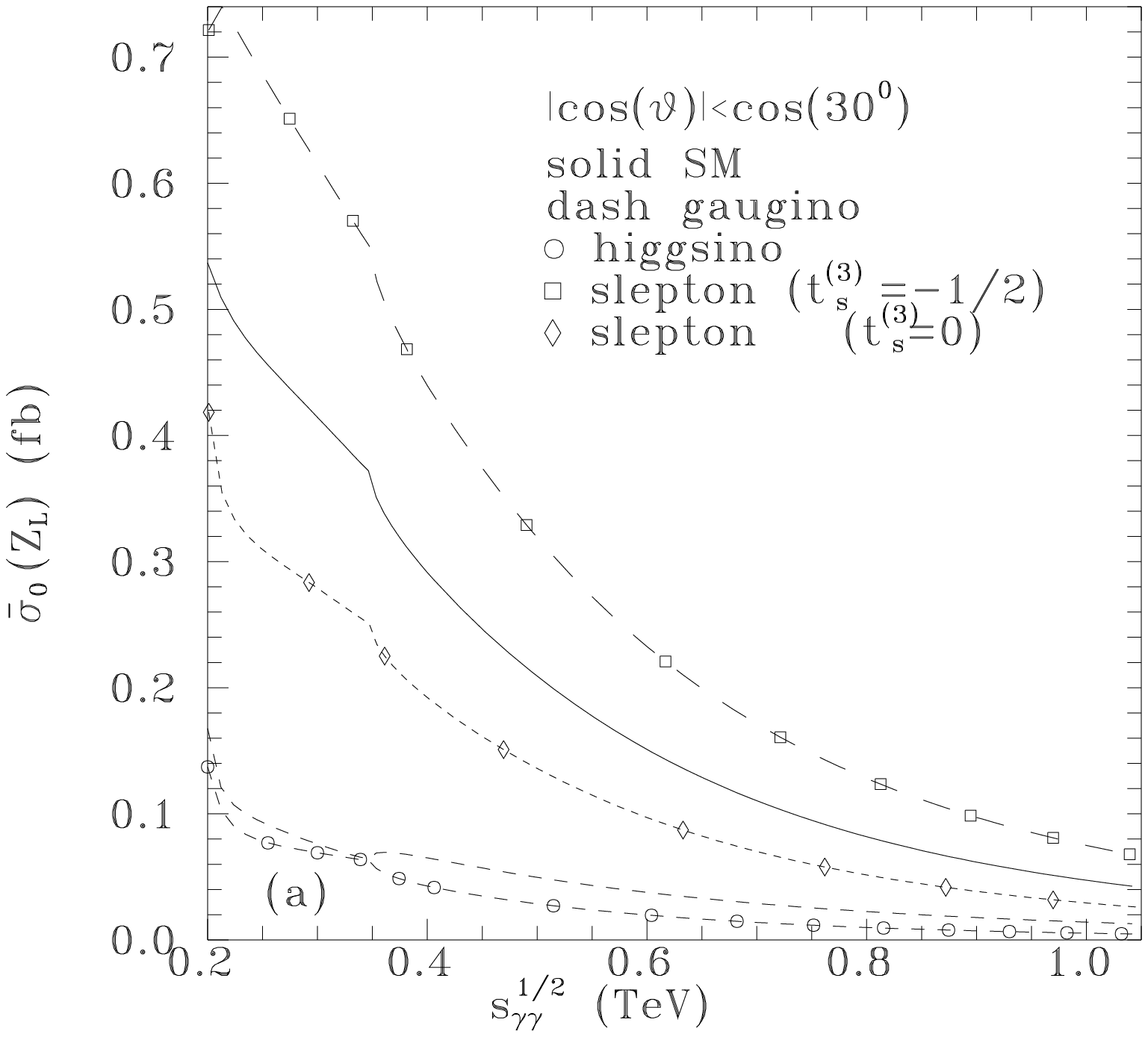,height=7.5cm}\hspace{0.5cm}
\epsfig{file=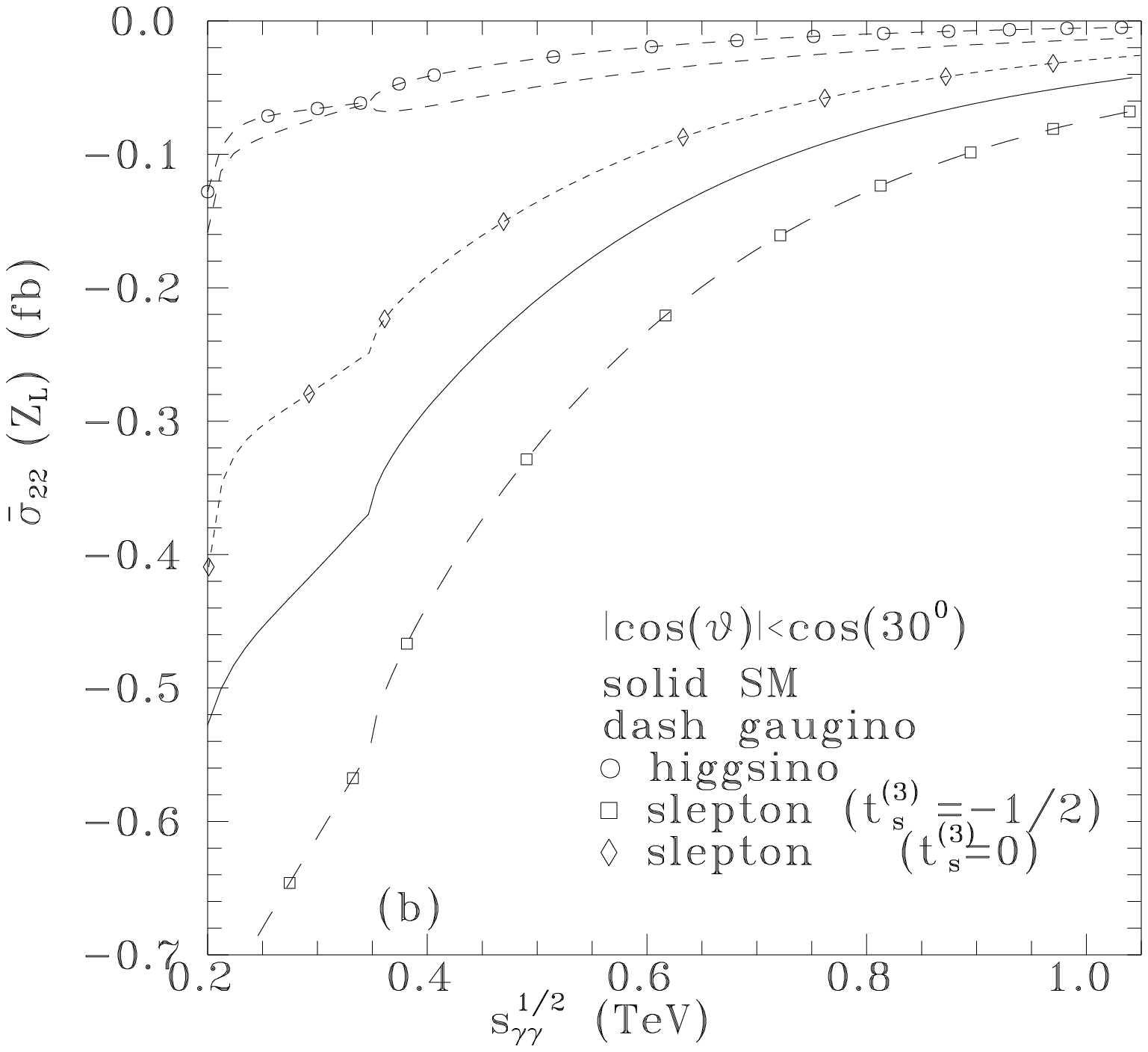,height=7.5cm}
\]
\vspace*{0.5cm}
\[
\epsfig{file=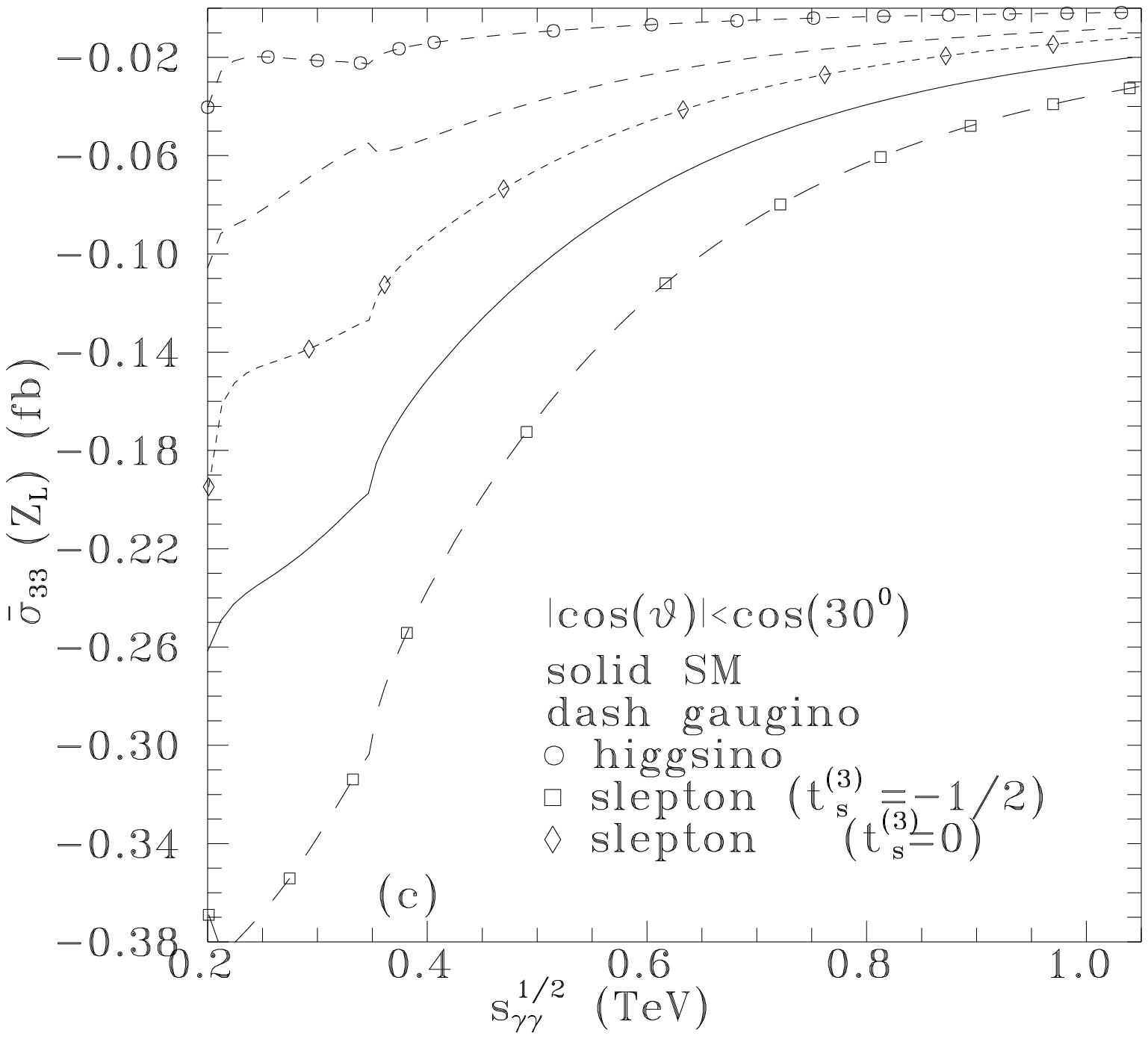,height=7.5cm}\hspace{0.5cm}
\]
\vspace*{0.5cm}
\caption[1]{Cross sections $\bar \sigma_0(Z_L)$, $\bar \sigma_{22}(Z_L)$,
and  $\bar \sigma_{33}(Z_L) $ for $Z_L$ production
in SM (solid) and in the presence
of a chargino or a charged slepton contribution denoted as
in Fig.\ref{sig}.}
\label{long}
\end{figure}

\begin{figure}[p]
\vspace*{-4cm}
\[
\epsfig{file=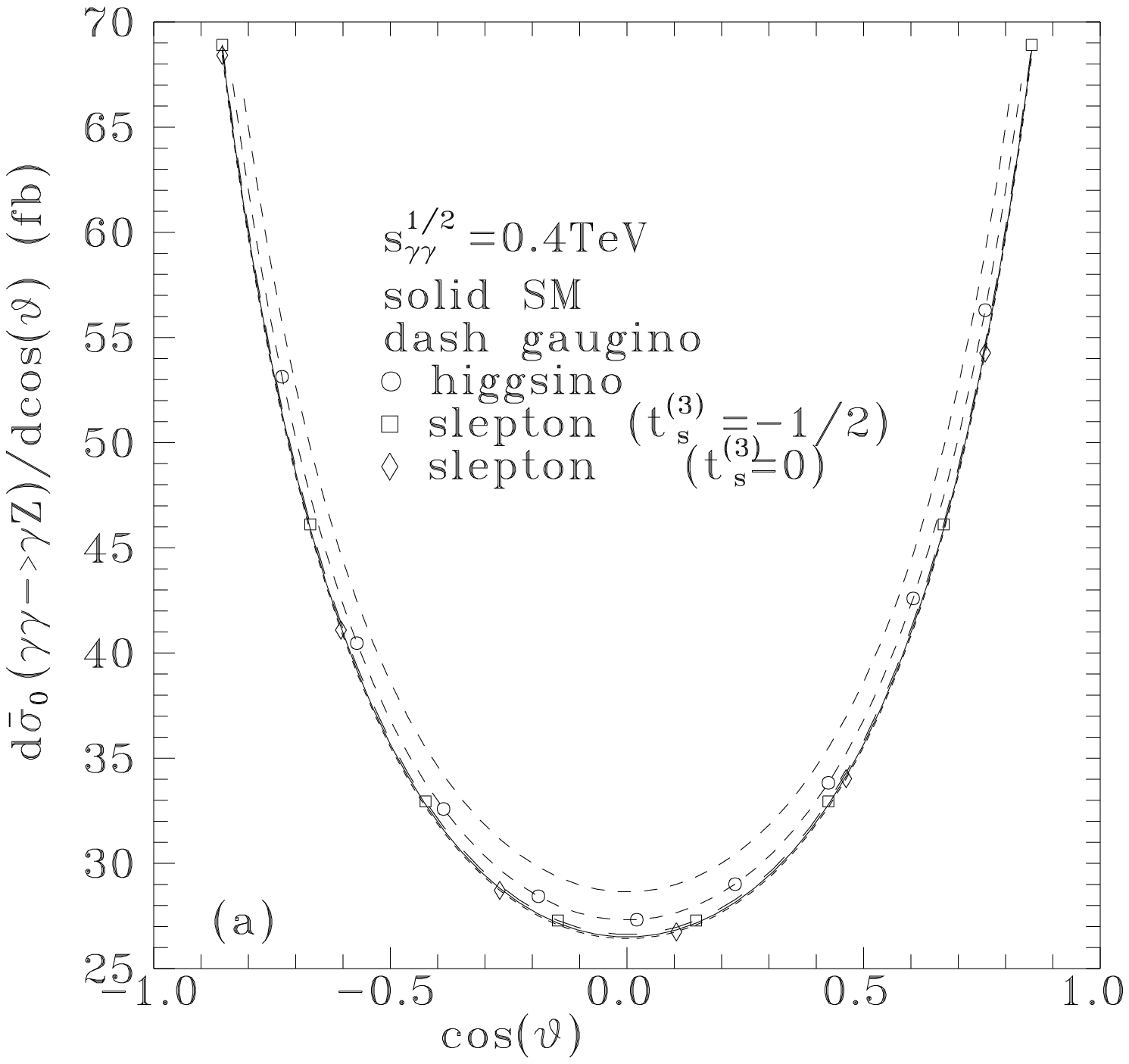,height=7.5cm}\hspace{0.5cm}
\epsfig{file=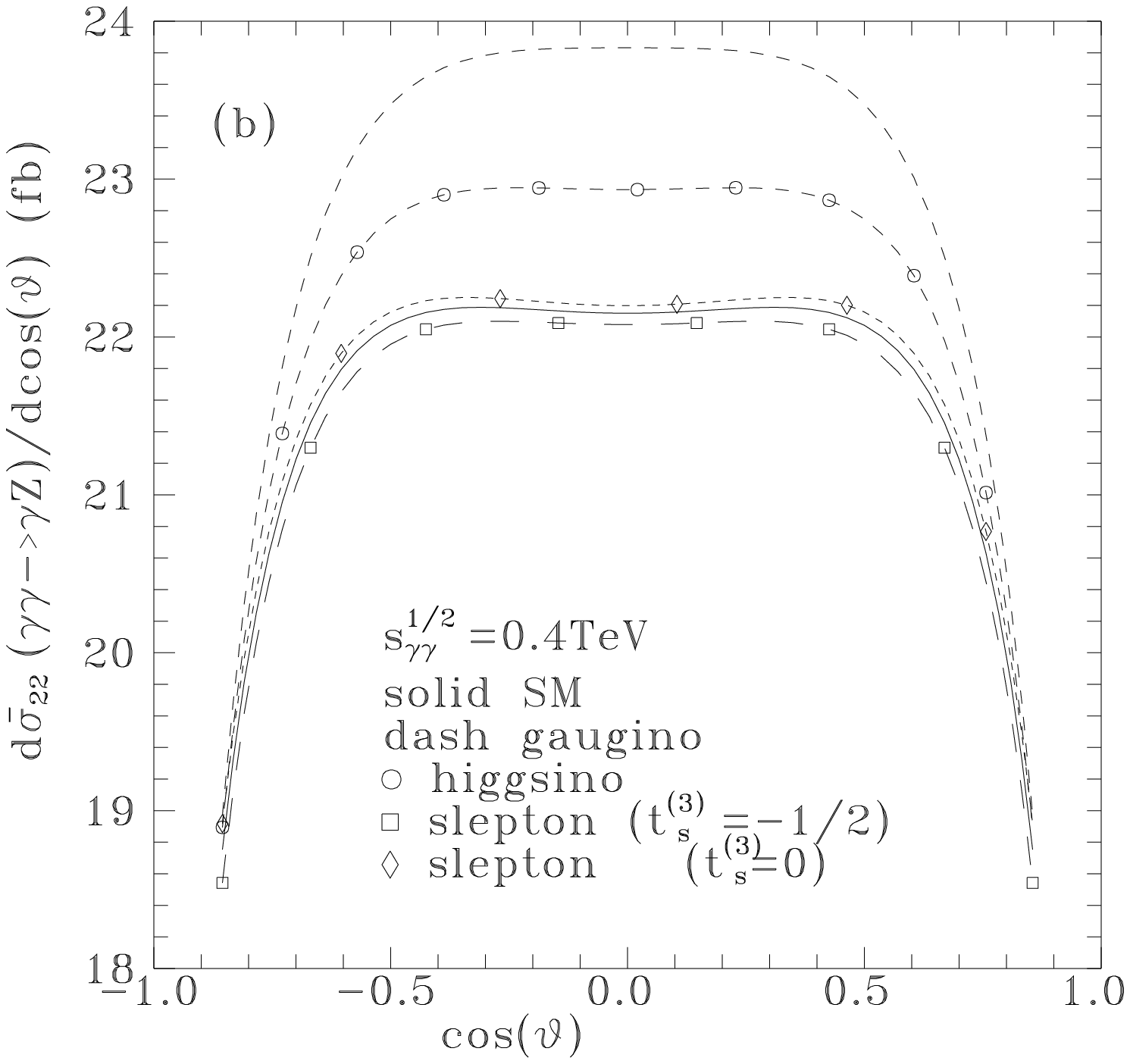,height=7.5cm}
\]
\vspace*{0.5cm}
\[
\epsfig{file=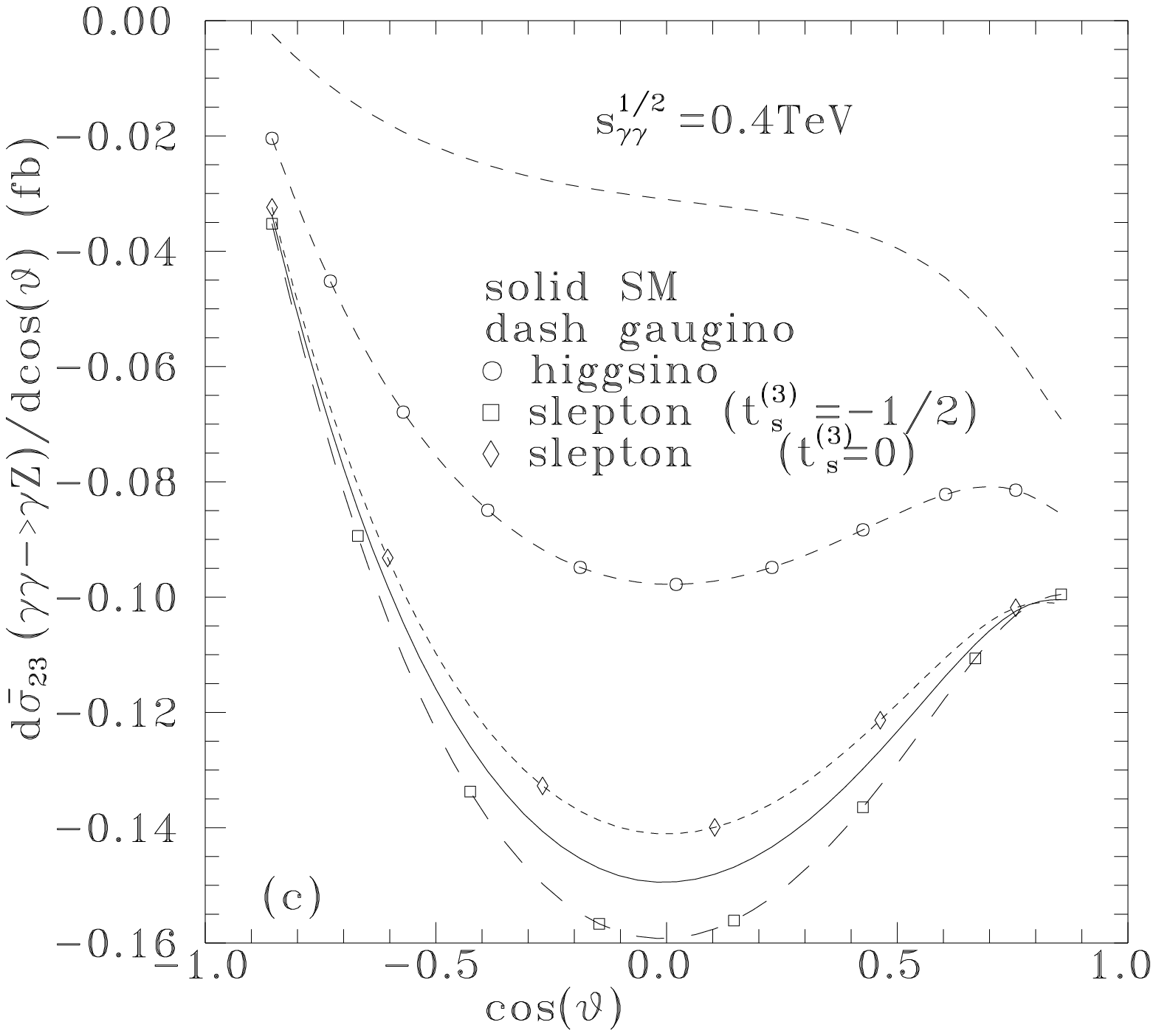,height=7.5cm}\hspace{0.5cm}
\epsfig{file=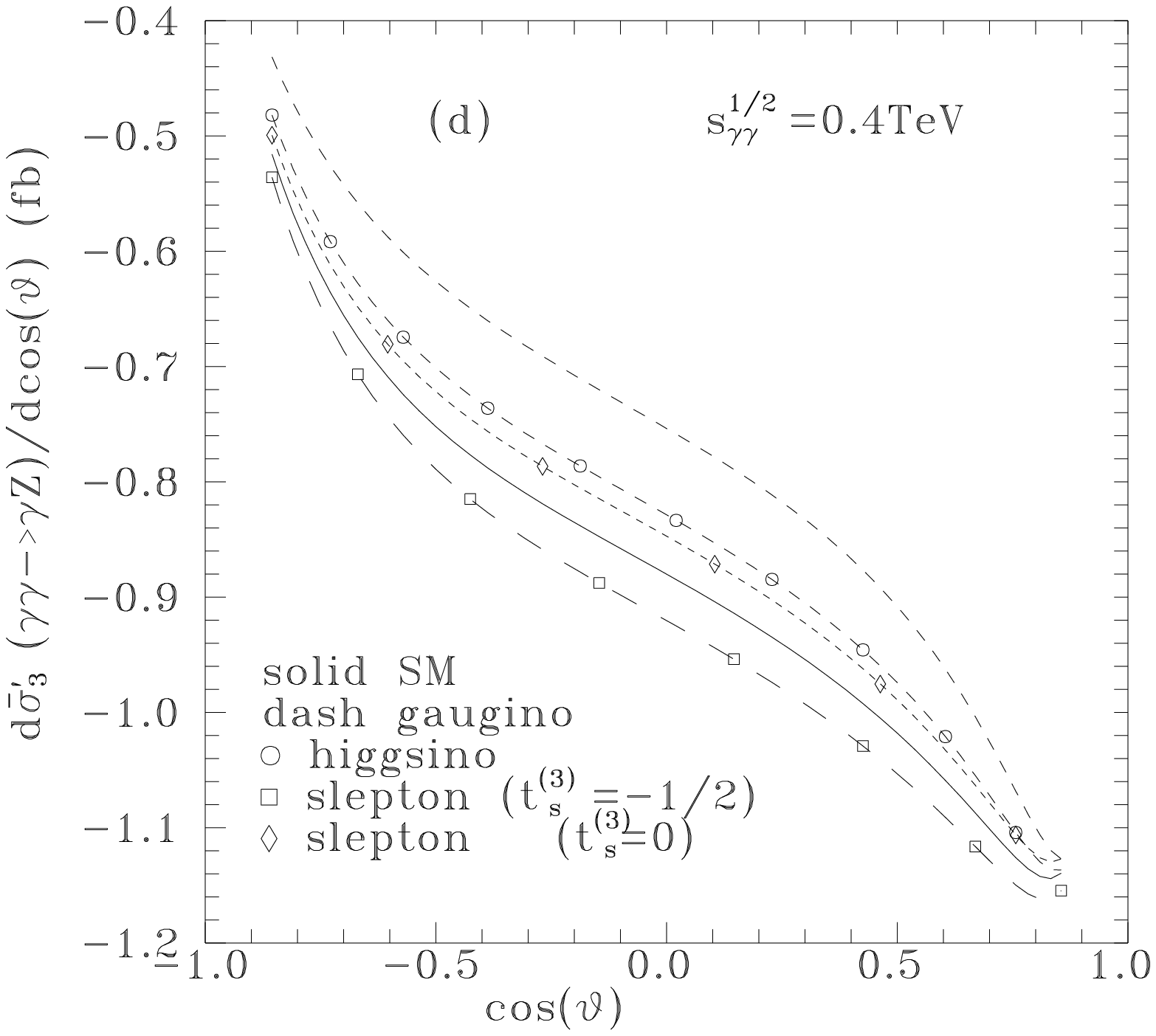,height=7.5cm}
\]
\vspace*{0.5cm}
\caption[1]{Angular distributions for
$d\bar{\sigma}_{0}/d\cos\vartheta^*$,
$d\bar{\sigma}_{22}/ d\cos\vartheta^*$,
$d\bar{\sigma}_{23}/d\cos\vartheta^*$,
$d\bar{\sigma}^\prime_{3}/d\cos\vartheta^*$.}
\label{ang}
\end{figure}

\clearpage
\newpage

\addtocounter{figure}{-1}

\begin{figure}[p]
\vspace*{-4cm}
\[
\epsfig{file=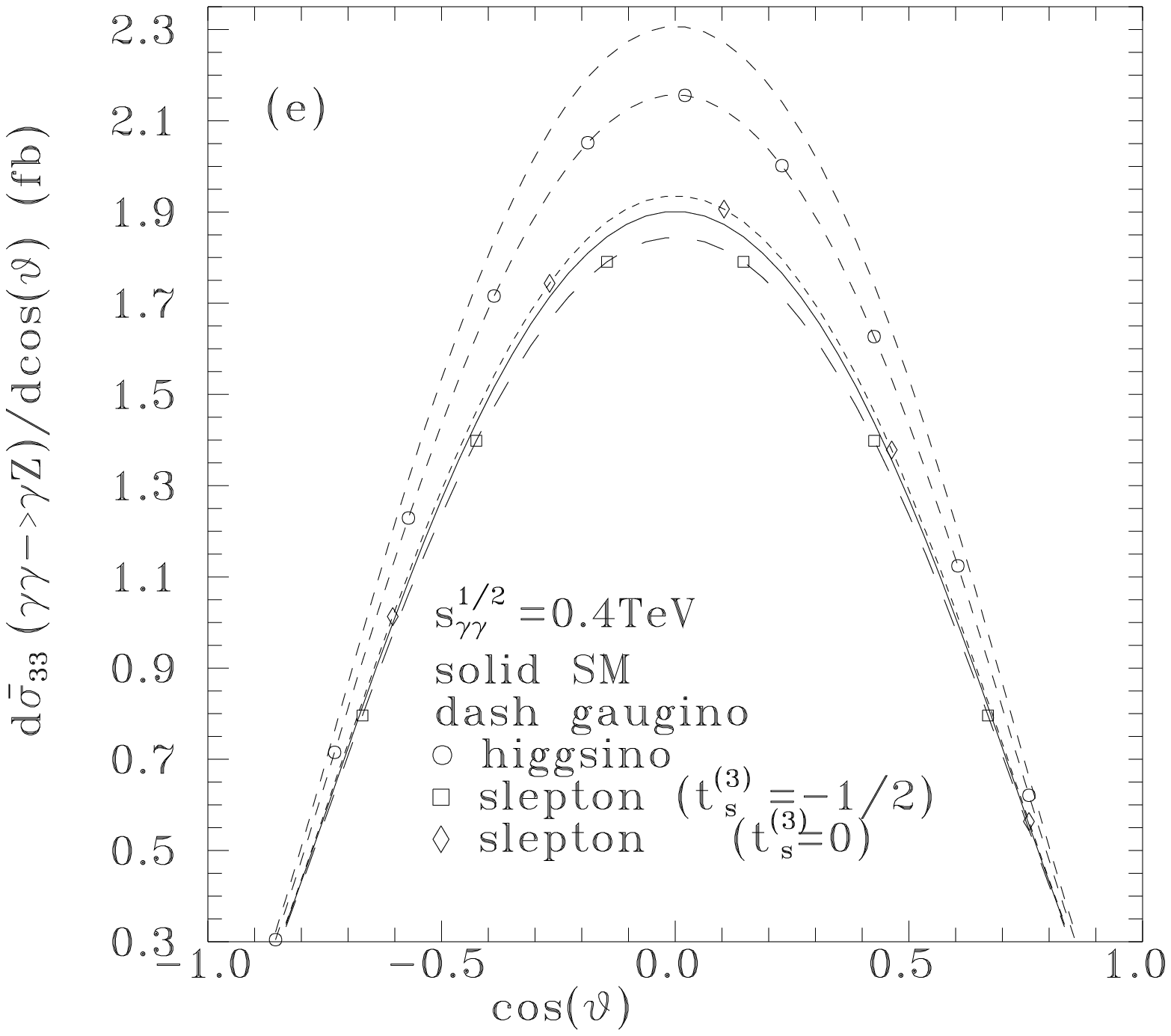,height=7.5cm}\hspace{0.5cm}
\epsfig{file=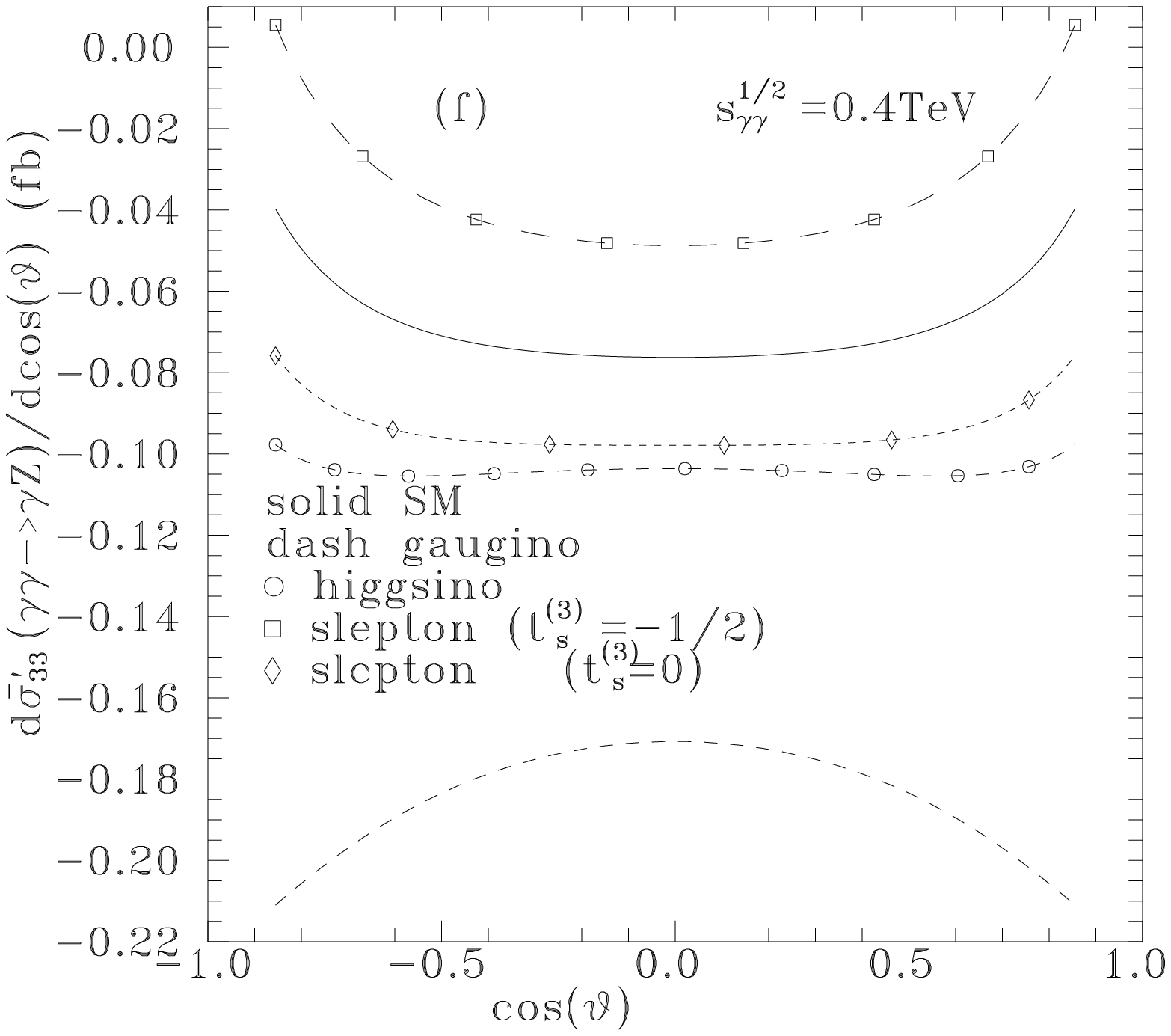,height=7.5cm}
\]
\vspace*{0.5cm}
\caption[1]{Angular distributions for
$d\bar{\sigma}_{33}/ d\cos\vartheta^*$,
$d\bar{\sigma}^\prime_{33}/ d\cos\vartheta^*$.}
\label{ang1}
\end{figure}

\end{document}